\documentclass[usenatbib]{mnras}

\usepackage{booktabs}
\usepackage{ae,aecompl}
\usepackage{graphicx}
\usepackage{amsmath}
\usepackage{amssymb}
\usepackage{multirow}
\usepackage{txfonts}
\usepackage[dvipsnames]{xcolor}

\newcommand{\keV}{{\rm keV}}

\newcommand{\Mpc}{{\rm Mpc}}
\newcommand{\GHz}{{\rm\, GHz}}

\newcommand{\beq}{\begin{equation}}   %

\newcommand{\eeq}{\end{equation}}   %

\newcommand{\beqa}{\begin{eqnarray}}   %

\newcommand{\eeqa}{\end{eqnarray}}   %

\newcommand{\beal}{\begin{align}}
\newcommand{\enal}{\end{align}}

\newcommand{\bspl}{\begin{split}}

\newcommand{\espl}{\end{split}}

\newcommand{\bsub}{\begin{subequations}}

\newcommand{\esub}{\end{subequations}}

\newcommand{\bmulti}{\begin{multline}}   %

\newcommand{\beqm}{\begin{mathletters}}   %

\newcommand{\eeqm}{\end{mathletters}}   %

\newcommand{\Te}{T_{\rm e}}

\newcommand{\vek} [1]{\mbox{\boldmath${#1}$\unboldmath}}

\newcommand{\pot}[2]{#1 \times 10^{#2}}

\newcommand{\Yp}{Y_{\rm p}}

\newcommand{\omb}{\Omega_{\rm b}h^2}
\newcommand{\omc}{\Omega_{\rm cdm}h^2}
\newcommand{\thetaMC}{\theta_{\rm MC}}

\newcommand{\tcmb}{T_{\rm CMB}}
\newcommand{\atwo}{A_{2\gamma}}

\newcommand{\Ho}{H_0}
\newcommand{\dIv}{\Delta I_{\,\nu}^{\rm \,rec}}
\newcommand{\dIvo}{\Delta I_{\,\nu,0}^{\rm \,rec}}

\newcommand{\Neff}{N_{\rm eff}}

\newcommand{\Ypb}{\Yp^{\rm BBN}}

\usepackage{xspace}
\newcommand{\FIRAS}{{\it COBE/FIRAS}\xspace}
\newcommand{\PIXIE}{{\it PIXIE}\xspace}
\newcommand{\SuperPIXIE}{{\it SuperPIXIE}\xspace}
\newcommand{\Voyage}{{\it Voyage~2050}\xspace}
\newcommand{\Voyagep}{{\it Voyage~2050+}\xspace}
\newcommand{\Voyagepp}{{\it Voyage~2050++}\xspace}
\newcommand{\Planck}{{\it Planck}\xspace}
\newcommand{\Millimetron}{{\it Millimetron}\xspace}

\newcommand{\dIvBBN}{\Delta I_{\,\nu}^{\rm \,rec, BBN}}
\newcommand{\NeffBBN}{\Neff^{\rm BBN}}

\newcommand{\ns}{{n_{\rm s}}}
\newcommand{\As}{{A_{\rm s}}}

\title[Forecasts for the CRR]{Sensitivity forecasts for the cosmological recombination radiation in the presence of foregrounds}

\author[Hart, Rotti and Chluba]{
Luke Hart$^{1}$\thanks{luke.hart@manchester.ac.uk}, Aditya Rotti$^{1}$\thanks{aditya.rotti@manchester.ac.uk} and Jens Chluba$^{1}$\thanks{jens.chluba@manchester.ac.uk}
\\
$^{1}$Jodrell Bank Centre for Astrophysics, Alan Turing Building, University of Manchester, Manchester M13 9PL \\}

\date{\vspace{-0mm} Accepted XXX. Received YYY; in original form ZZZ}

\pubyear{2020}

\begin{document}
\label{firstpage}
\pagerange{\pageref{firstpage}--\pageref{lastpage}}
\maketitle

\begin{abstract}
The cosmological recombination radiation (CRR) is one of the inevitable $\Lambda$CDM spectral distortions of the cosmic microwave background (CMB). While it shows a rich spectral structure across dm-mm wavelengths, it is also one of the smallest signals to target. Here we carry out a detailed forecast for the expected sensitivity levels required to not only detect but also extract cosmological information from the CRR in the presence of foregrounds. We use {\tt CosmoSpec} to compute the CRR including all important radiative transfer effects and modifications to the recombination dynamics.
We confirm that detections of the overall CRR signal are possible with spectrometer concepts like $\SuperPIXIE$. However, for a real exploitation of the cosmological information, a $\simeq 50$ times more sensitive spectrometer is required. While extremely futuristic, this could provide independent constraints on the primordial helium abundance, $\Yp$, and probe the presence of extra relativistic degrees of freedom during BBN and recombination.
Significantly improving the constraints on other cosmological parameters requires even higher sensitivity (another factor of $\simeq 5$) when considering a combination of a CMB spectrometer with existing CMB data. To a large part this is due to astrophysical foregrounds which interestingly do not degrade the constraints on $\Yp$ and $\Neff$ as much. A future CMB spectrometer could thus open a novel way of probing non-standard BBN scenarios, dark radiation and sterile neutrinos.
In addition, inflation physics could be indirectly probed using the CRR in combination with existing and forthcoming CMB anisotropy data.
\end{abstract}


\section{Introduction}\label{sec:intro}
Advances in studies of the cosmic microwave background (CMB) temperature and polarisation anisotropies led to unprecedented steps forward in determining fundamental parameters of our Universe \citep{COBE4yr,wmap9params,Planck2015params,Planck2018over}. However, anisotropies are not the only source of information from the CMB: we can also measure its energy distribution. 
The first constraints on the CMB energy spectrum were obtained by \FIRAS \citep{Mather1994, Fixsen1996,Fixsen2002}. 
These measurements confirmed the blackbody nature of the average CMB spectrum, $I_\nu$, to high precision, establishing one of the fundamental pillars of modern cosmology.
Departures of the CMB spectrum from a blackbody, so-called spectral distortions, were limited to $|\Delta I_{\nu}/I_{\nu}|\lesssim 10^{-5}-10^{-4}$. This provides tight constraints on the thermal history of our Universe, ruling out cosmologies with significant energy release at redshift $z\lesssim {\rm few}\times10^6$ \citep{Mather1994, Fixsen1996, Fixsen2011}. 

The theory of distortions has been studied extensively in previous literature \citep[e.g.,][]{Zeldovich1969, Sunyaev1970mu, Zeldovich1972, Illarionov1974, Danese1981, Burigana1991, Hu1993, Chluba2011therm}. 
One can broadly categorize spectral distortions as $\mu$- and $y$-type distortions \citep{Sunyaev1970mu, Zeldovich1969}. At $z\gtrsim\pot{2}{6}$, the cosmic radiation is quickly thermalised; Compton scattering, double Compton and Bremsstrahlung emission efficiently redistribute and produce photons to form a blackbody spectrum. These processes slow down at lower redshifts and for energy injection at $5\times 10^4 \lesssim z \lesssim \pot{2}{6}$, $\mu$-type distortions are formed. At even later times, when also Compton scattering becomes inefficient, $y$-type distortions are created. In the past years, space missions such as \PIXIE and enhanced versions have been considered with the capability of measuring CMB spectral distortions at a spectral sensitivity that is more than $\simeq 10^3$ times higher than that of \FIRAS \citep{Kogut2011PIXIE, Kogut2016SPIE, PRISM2013WPII, Kogut2019BAAS}. This could allow us to probe a wide range of standard and non-standard physics \citep[see][for broad overview]{Chluba2019BAAS, Chluba2019Voyage}.

One of the inevitable spectral distortions predicted in $\Lambda$CDM is the \emph{cosmological recombination radiation} (CRR) \citep[for review see][]{Sunyaev2009}. 
In contrast to the $\mu$- and $y$-distortions, this signal has many spectral features and is caused by the injection of photons at late stages, when thermalization processes are inefficient\footnote{For a broad discussion of photon injection distortions see \citet{Chluba2015GreensII}.}.
Historically, the recombination process was first studied using effective 3-level atom calculations, with a specific focus on the crucial HI Lyman-$\alpha$ distortion and 2s-1s two-photon decay process \citep{Zeldovich68, Peebles68}. These calculations established that cosmological recombination occurs as an out-of-equilibrium process, with radiative transitions controlling the recombination dynamics.
Subsequently, it was argued that the recombination lines from excited levels of hydrogen produce additional photons at dm-mm wavelengths, with many spectral features expected in the CMB spectrum \citep{Dubrovich1975}.

While the earliest recombination calculations already allowed anticipating the importance and beauty of the recombination signal as a potential cosmological probe, many computations followed to precisely model the CRR \citep{RybickiDell94,DubroVlad95,Dubrovich1997, Burgin2003, Kholu2005, Wong2006, Jose2006, Chluba2006b, Chluba2007, Jose2008}.  
The early computations were in particular limited by the availability of extensive atomic databases for hydrogen and helium. In addition, the modeling of radiative transfer effects was computationally challenging. 
These problems were only overcome once significant activity started to obtain accurate computations of the cosmological recombination history in preparation for the analysis of CMB data from \Planck \citep{Chluba2006, Jose2006, Kholu2006, Chluba2006b, Chluba2007, Kholupenko2007, Switzer2007I, Switzer2007II, Switzer2007III, Jose2008, Chluba2008a, Hirata2008, Chluba2008b, Chluba2009, Grin2009, Chluba2009c, Jose2010, Chluba2010, Yacine2010b, Yacine2010, Shaw2011}, leading to the development of the accurate recombination codes {\tt CosmoRec} \citep{Chluba2010b} and {\tt HyRec} \citep{Yacine2010c}.

\begin{figure}
    \centering
    \includegraphics[width=\columnwidth]{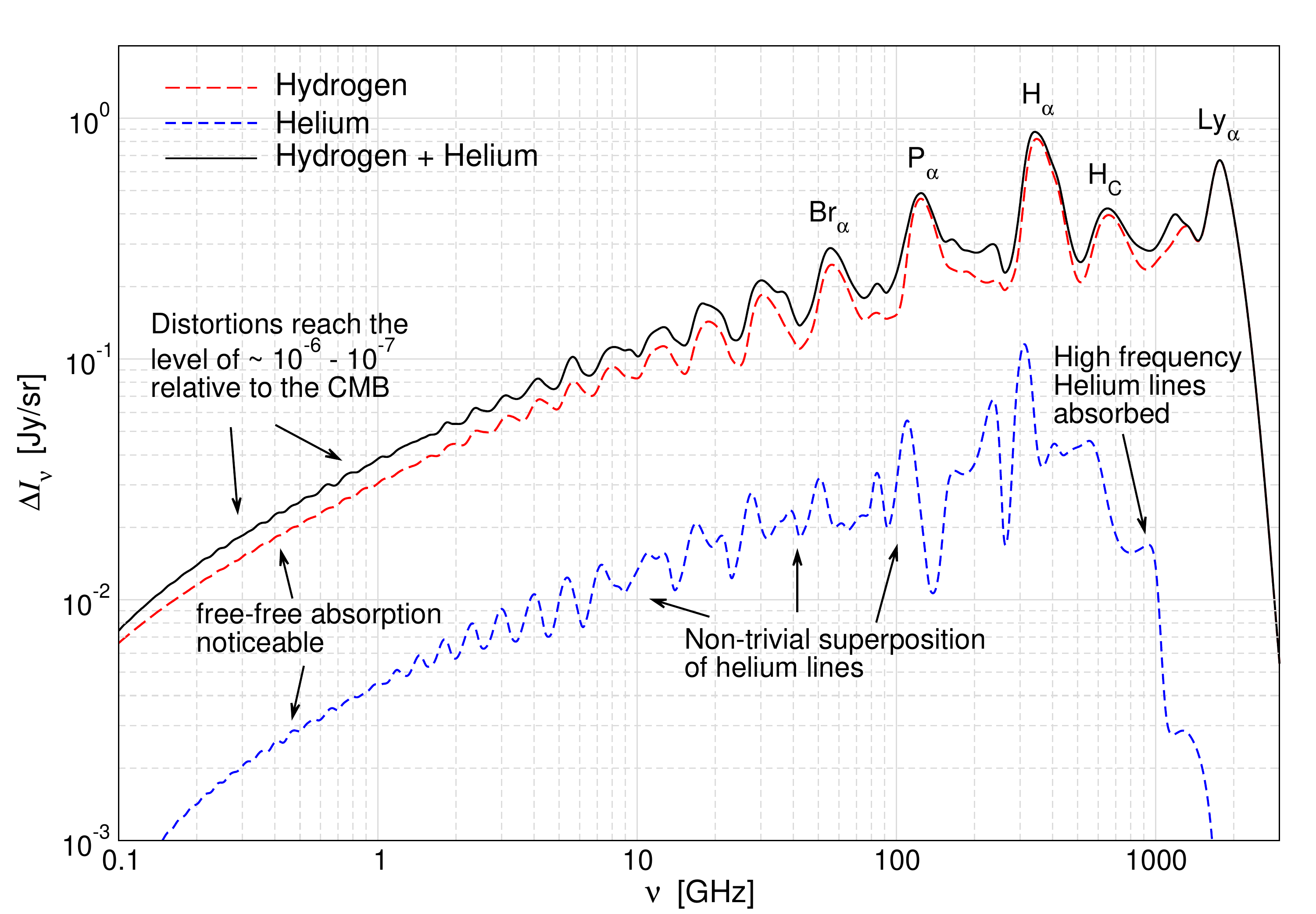}
    \vspace{-3mm}
    \caption{State of the art computation of the CRR using {\tt CosmoSpec}. The spectral series of hydrogen ($z\simeq 1400$) dominate the total emission in the CRR ({\it dashed, red line}). The helium distortion ({\it dashed, blue line}) is the net accumulation of the HeI and HeII emission from $z\simeq 2000$ and $z\simeq 6000$, respectively. The unique spectral shapes given by the CRR would provide us with a revolutionary new cosmological probe of the atomic physics in the early Universe. The figure was adapted from \citet{Chluba2019Voyage}.}
    \label{fig:CRR}
 \vspace{-3mm}
\end{figure}

For detailed computations of the CRR, one important aspect is that the populations in the angular momentum substates have to be distinguished \citep{Jose2006, Chluba2007}. This strongly affects the amplitude and shape of the CRR at all frequencies, but also the recombination dynamics at late times, which was previously omitted.  
An additional important step towards accurate computations of the CRR was the inclusion of the free-bound emission \citep{Chluba2006b}. In total, the free-bound continua lead to the production of one photon per recombined HI atom, with the most visible feature caused by the HI Balmer-continuum around $\nu\simeq 600-700\,\GHz$ (see Fig.~\ref{fig:CRR}), which was entirely omitted before. Aside from more detailed treatments of the recombination dynamics and the inclusion of more atomic shells, these two refinements led to the largest changes to the HI recombination radiation in comparison to earlier works.

The total recombination radiation also includes the emission from neutral helium and singly-ionized helium (see Fig.~\ref{fig:CRR}), with fine and hyperfine structure transitions playing crucial roles \citep{Jose2008}. It is very important to model the acceleration of helium recombination by the absorption of photons in the HI Lyman-continuum \citep{Kholupenko2007, Switzer2007I}. This was shown to lead to a sharpening of the lines from neutral helium, providing a direct tracer of the recombination dynamics at $z\simeq 2000$ \citep{Jose2008}. Additional important effects on the helium recombination radiation stem from feedback processes \citep{Chluba2009c, Chluba2012HeRec} and electron scattering \citep{Jose2008}, which modify the shape of the recombination features and have to be included carefully, e.g., when constraining the helium abundance.

For the forecasts presented here, we will use {\tt CosmoSpec} \citep{Chluba2016CosmoSpec}, which is based on an extended version of {\tt CosmoRec} \citep{Chluba2010b} and includes all of the aforementioned effects, delivering the most detailed and accurate computations of the CRR to date. To accelerate the calculation, {\tt CosmoSpec} uses an effective conductance method \citep{Yacine2013RecSpec}, which was extended to the modeling of neutral and singly-ionized helium, for the first time including all free-bound and bound-bound transitions. Furthermore, detailed computation of the neutral and singly-ionized helium recombination process is performed, including all important radiative transfer and feedback effects between the atomic species \citep{Chluba2016CosmoSpec}. 

The first obvious question about the CRR is its {\it detectability}. The CRR is a small signal in the context of spectral distortions, peaking at $\Delta I_\nu \simeq 0.9$~Jy/sr (Fig.~\ref{fig:CRR}), some 3 orders of magnitude below the average Compton-$y$ distortion expected from all the clusters in the Universe \citep[e.g.,][]{Refregier2000, Hill2015}. It is furthermore obscured by intense galactic and extra-galactic foregrounds, which have to be included for reliable forecasts.
However, since the CRR exhibits significant frequency-dependence one can look for unique $\simeq {\rm 10-30 \,nK}$ variations in the broad CMB spectrum when extracting this signal \citep{Sunyaev2009}.

Two main avenues for the CRR detection have been considered: measurements at low frequencies ($\nu \simeq 3\,\GHz-6\,\GHz$) using arrays of radiometers from the ground \citep{Mayuri2015}, and CMB spectroscopy with future space experiments \citep{Vince2015}. These works showed that one of the main challenges is foreground removal. For an all-sky space mission, sensitivities at the level of $\simeq 0.25$~Jy/sr covering $\nu \simeq 30\GHz-3{\rm THz}$ with frequency resolution $\Delta \nu\simeq 15\,\GHz$ were shown to allow a $\simeq 5\sigma$ detection of the CRR \citep{Vince2015}. This is $\simeq 10$ times more sensitive than \PIXIE \citep{Kogut2011PIXIE, Kogut2016SPIE}, but otherwise has very similar specifications.\footnote{Details for the various experimental concepts will be given in Sect.~\ref{sec:experiments}, but for reference we quote the sensitivity at frequency $\nu=100\,\GHz$, which for a 4-year \PIXIE mission results in $\sigma(I_\nu)\simeq 2.5$~Jy/sr.} 
Another analysis, following the refined foreground and primordial distortion modeling of \citet{Abitbol2017}, reached very similar conclusions, showing that a $\simeq 2\sigma$ detection of the CRR could be possible with concepts like \SuperPIXIE \citep{Kogut2019BAAS}, while an $\simeq 8\sigma$ detection could come into reach for an even more ambitious \Voyage experiment \citep{Chluba2019Voyage}.
Respectively, these two concepts can be thought of as $\simeq 6$ and $\simeq 30$ times more sensitive than \PIXIE, corresponding to sensitivities at the level of $\simeq 0.4$~Jy/sr and $\simeq 0.08$~Jy/sr, respectively. One of the goals of this paper is to provide the details of the \SuperPIXIE and \Voyage forecasts.

While expected to be extremely challenging, the next big question is {\it what and how can we learn from detailed measurements of the CRR and which cosmological parameters it will help to constrain?}  
It is clear that the CRR depends directly on parameters such as the baryon density, CMB monopole temperature and helium abundance. First detailed demonstrations based on refined recombination calculations were given in \citet{Chluba2008T0}. There it was shown that the monopole temperature, $\tcmb$, adjusts the time of recombination and therefore the recombination line positions. In addition, the overall amplitude of the CRR was shown to be directly proportional to the baryon density, $\Delta I^{\rm rec}_{\nu}\propto \omb$. On the other hand, the value of the present-day Hubble parameter, $H_0$, has a more minor effect on the CRR \citep{Chluba2008T0}, although the response depends on which quantities are kept fixed (Sect.~\ref{sec:standard}). Similarly, the dependence on $\omc$ is expected to be small, as it only indirectly affects the CRR through the expansion rate.
All these statements can be understood when considering the physics of recombination (even in the earliest computations) and were directly confirmed with {\tt CosmoSpec}.

One of the unique opportunities with the CRR is a measurement of the primordial helium abundance, parametrized through $\Yp$. The CRR provides one of the cleanest probes in this respect, as it directly depends on the unreprocessed, pristine helium abundance, long before the first stars even formed \citep{Chluba2008T0}. With {\tt CosmoSpec} it now has become possible to accurately propagate the $\Yp$ dependence to the CRR.
Additional opportunities lie in the possibility to break degeneracies between $\Yp$ and the effective number of relativistic degrees of freedom, $\Neff$, using the CRR \citep{Chluba2016CosmoSpec}. The CRR could furthermore allow probing non-standard physics, such as {\it pre-recombinational energy} release \citep{Chluba2008c} or specifically \emph{dark matter annihilation} and \emph{decay} \citep{Chluba2010a}, and the \emph{variation of fundamental constants} \citep{Chluba2016CosmoSpec, Hart2017}. 

An initial, simple estimate for the constraining power of the CRR on the standard parameters was presented in \citet{Fendt2009Thesis}, based on the computations of \citet{Jose2006, Chluba2007, Jose2008}. The analysis interpolated pre-tabulated outputs of the CRR used for the development of {\tt RICO} \citep{Fendt2009} to obtain these estimates. This did not include a detailed treatment of the helium recombination spectrum and related feedback processes, nor a modeling of foregrounds; however, it was shown that precise measurements of $\Omega_{\rm b}h^2$ and $\Yp$ can be obtained if a $\simeq 1\%$ fractional sensitivity on the total CRR (i.e., variable sensitivity of $\simeq 10^{-3}-10^{-2}$~Jy/sr or more than 100 times more sensitive than \PIXIE) can be reached. In this case, the constraints on these parameters respectively improved by a factor of $\simeq 10$ and $\simeq 6$ when combined with \Planck.\footnote{For their estimate a simple Gaussian likelihood based on the \Planck bluebook noise level was used.} 
As similar analysis was recently performed by \citet{Sarkar2020}. 
However, in both studies, the conversion of sensitivities to compare with \PIXIE and its enhanced versions seem problematic and, probably more importantly, the degradation from foreground signals was not included, further necessitating a more careful study with {\tt CosmoSpec}.

Here, we focus on forecasts of the main CRR parameters assuming a standard cosmological background and also extensions that treat $\Yp$ and $\Neff$ as independent variables. Parameters like the Thomson optical depth, $\tau$, or curvature power spectrum parameters ($\ns$ and $\As$) do not enter the recombination problem, and thus are not considered directly.
The CMB monopole temperature, $\tcmb$, can be precisely fixed by measuring the blackbody part of the CMB spectrum itself, such that for the CRR the most promising cosmological parameters are $\omb$, $\omc$ and $\Yp$, given that the dependence on $h$ and $\Neff$ is known to be small \citep{Chluba2008T0, Chluba2016CosmoSpec}. 
However, it turns out that $\omc$ also has a rather small effect on the CRR and thus is not easily constrained. In addition, the marginalization of foregrounds strongly hampers the constraints on $\omb$, leaving $\Yp$ as the most interesting target. Further including Big Bang Nucleosynthesis (BBN) relations \citep[e.g.,][]{Steigman2007, Iocco2009}, we can in principle also obtain measurements of $\Neff$ at BBN through the effect on $\Yp$, as we illustrate here.
This implies that the CRR could become a powerful new probe of non-standard BBN scenarios, dark radiation and sterile neutrinos.
Indirectly, one can also improve on other cosmological parameters like $\ns$ and $\As$ when combining with existing CMB anisotropy data from \Planck \citep{Planck2018params}, however, the required sensitivities are extremely futuristic.

The paper is structured as follows:
In Sect.~\ref{sec:cosmology}, we will first illustrate the main parameter dependence of the CRR. As one non-standard parameter, we freely vary the 2s-1s two-photon decay rate to illustrate the potential of distortion constraints on recombination dynamics and physics.
Our forecasts are based on simple Fisher estimates. The details are presented in Sect.~\ref{sec:measurements} together with some of the experimental parameters. We will consider separate distortion constraints from spectrometer concepts like \SuperPIXIE, \Voyage and even enhanced versions. Recent works relating to \Millimetron\footnote{\url{www.millimetron.ru/}} suggest that significantly higher sensitivities using foldable mirrors in space might become available in the future. In addition, ideas of going to the moon \citep{Silk2018Nature} could improve our observational capabilities many-fold, such that these futuristic estimates remain interesting. 
In Sect.~\ref{sec:foregrounds} and~\ref{sec:other_SD}, we present a brief recap of the main foregrounds and other distortion signals that need to be marginalized over for reliable forecasts.

In Sect.~\ref{sec:forecasts}, we then summarize the results of our forecasts. We present spectrometer-only constraints as well as in combination with existing CMB data. To illustrate the importance of foregrounds we perform runs with varying levels of foreground complexities. Ultimately we show that a spectrometer with ten times the sensitivity of \Voyage in principle could obtain interesting constrains on parameters like $\Yp$ and $\Neff$, while improvements on the other parameters require even higher sensitivity.

\vspace{0mm}
\section{Cosmology with recombination lines}
\label{sec:cosmology}
The CRR is a superposition of hydrogen ($z_{\rm HI}\simeq 1400$) and helium ($z_{\rm HeI}\simeq 2200$ and $z_{\rm HeII}\simeq 6000$) recombination line emission, which leads to unique spectral features that allow us to constrain various cosmological parameters and epochs \citep{Sunyaev2009}. 
While with {\tt CosmoSpec} \citep{Chluba2016CosmoSpec} it is possible to accurately compute the CRR given a set of standard cosmological parameters, $\vek{p}$, for the Fisher forecasts presented below, it is sufficient to model the spectrum as a perturbation around the fiducial $\Lambda$CDM values.
In this case, the dependence of the CRR intensity, $\dIv$, at frequency $\nu$ can be obtained using
$\dIv \approx \dIvo+\sum_{i}\left.\partial\dIv/\partial p_i \right|_{\,\vek{p} = \vek{p}_0}\,\Delta p_i$.
Here, $\dIvo$ is the recombination radiation with a fiducial $\Lambda$CDM cosmology represented by $\vek{p}_0$ (see Fig.~\ref{fig:CRR}).
Non-linear corrections will become important close to the detection limits, slightly hampering the forecasts of the spectrometer-only constraints; however, once external information is added, this no longer is an issue.
In this study, the fiducial cosmology is represented by the most recent \emph{Planck} results \citep{Planck2018params}. For the present-day CMB temperature we use $\tcmb=2.7255$ \citep{Fixsen1996, Fixsen2009}. We furthermore assume a flat Universe.

\begin{figure}
    \centering
    \includegraphics[width=\columnwidth]{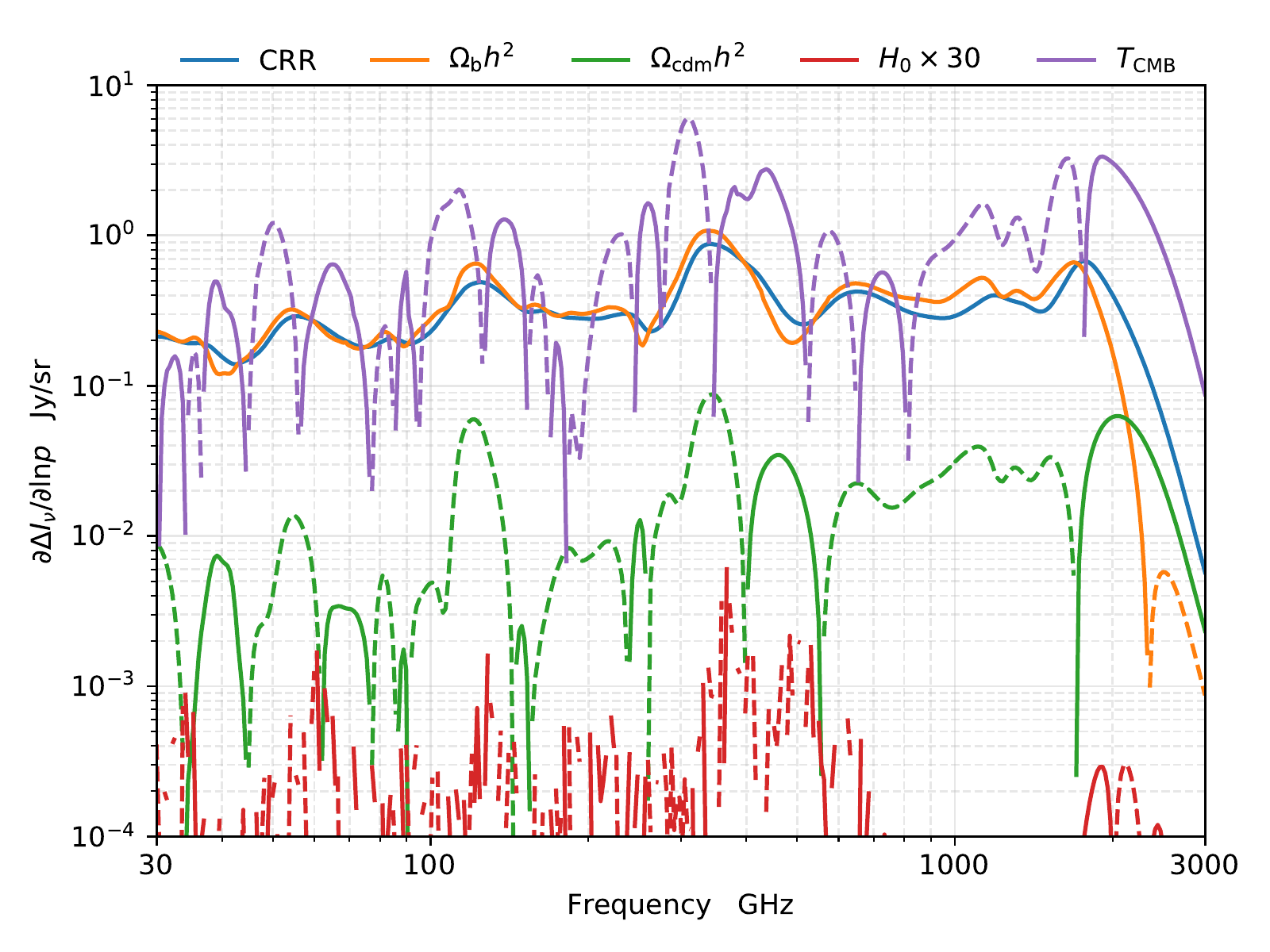}
    \\[-0mm]
    \includegraphics[width=\columnwidth]{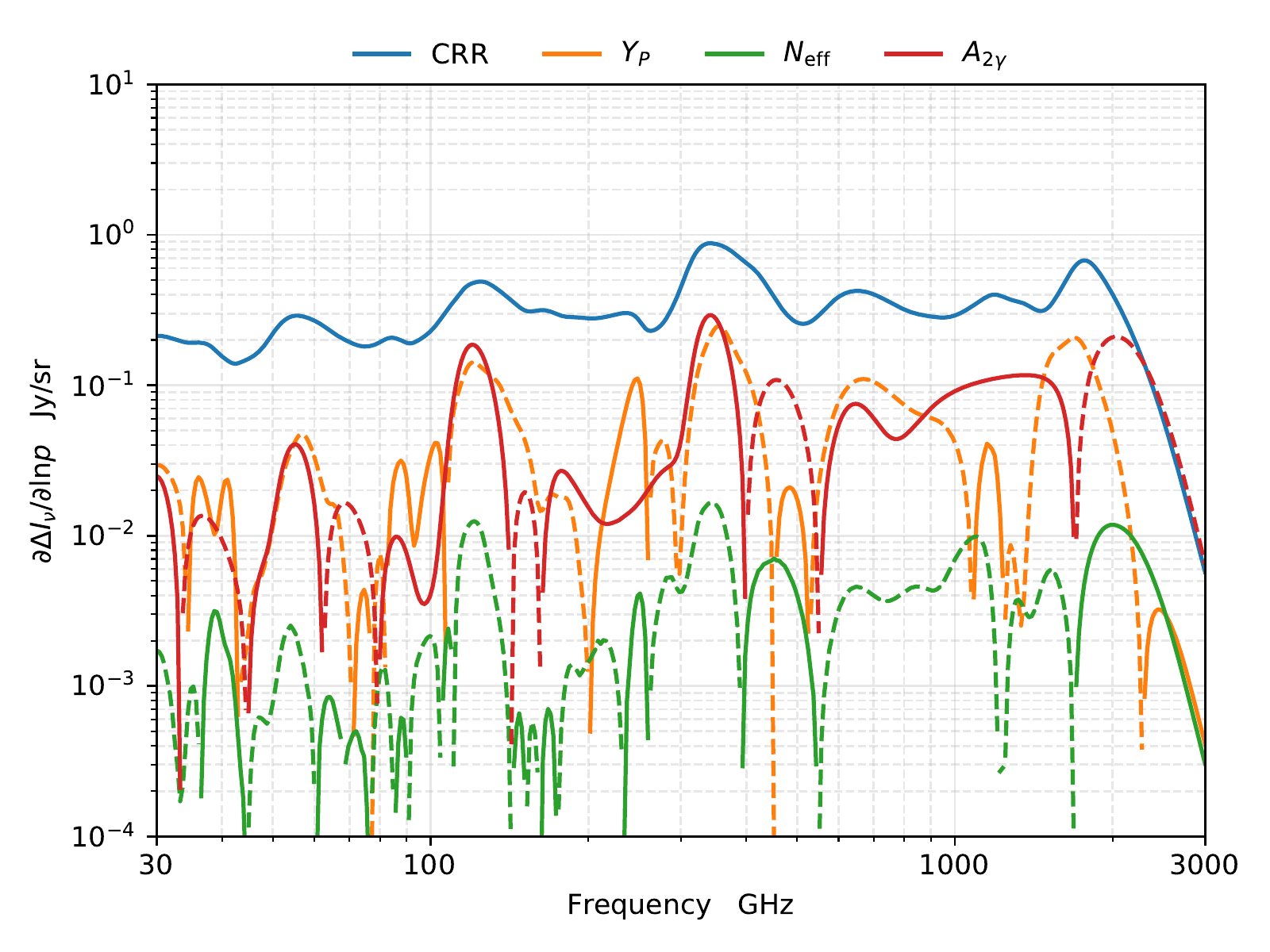}
    \caption{Variations in the recombination lines for different cosmological parameters. \emph{Top}: Standard $\Lambda$CDM parameter derivatives. \emph{Bottom}: Simple parametric extensions of the $\Lambda$CDM parameters. The \PIXIE frequency range has been used in these figures \citep{Kogut2016SPIE}. Dashed line parts are negative branches. Note also that we rescaled the response for $\Ho$.}
    \label{fig:parameters}
\end{figure}

We now first look at the standard $\Lambda$CDM parameters relevant to the CRR: the baryon density $\omb$, the cold dark matter density $\omc$, the Hubble constant $\Ho$ and the CMB monopole temperature $\tcmb$.
We then study extended parameters: the primordial helium abundance, $\Yp$, the effective number of neutrino species, $\Neff$, and the two-photon decay rate for hydrogen, $\atwo$, to illustrate future possibilities with the CRR.
All the final responses are summarized in Fig.~\ref{fig:parameters}, where we show the logarithmic derivative $\partial\dIv/\partial\ln p$. This allows a better comparison of the effective parameter responses relative to the full CRR (dotted blue lines in Fig.~\ref{fig:parameters}) and among each other.

\vspace{-0mm}
\subsection{Standard cosmological parameters}
\label{sec:standard}
One of the largest effects arises from changes in $\omb$, which determines the total number of hydrogen and helium atoms and thus the overall amplitude of the CRR, i.e., $\dIv\propto\omb$ \citep{Chluba2008T0, Chluba2016CosmoSpec}. From this simple relation one expects $\partial\dIv/\partial \ln\omb \propto \dIv$. To leading order this is indeed clearly visible in Fig.~\ref{fig:parameters}. However, since $\omb$ also directly affects the recombination dynamics, small departures in the shape are visible. To obtain the derivative with respect to $\omb$, we kept $\Yp$ fixed, but will return to this aspect below (Sect.~\ref{sec:BBN}).

Another large effect on the CRR is expected from variations of $\tcmb$, as it directly affects the position of the lines \citep{Chluba2008T0}. From this simple understanding one can anticipate $\partial\dIv/\partial \ln\tcmb \propto \partial \dIv/\partial \ln\nu$. Looking at the result in Fig.~\ref{fig:parameters} broadly confirms this statement, with extrema in the CRR coinciding with the nulls of $\partial\dIv/\partial \ln\tcmb$ and largest variations visible close to the flanks of the recombination line features.
The relative variation due to $\tcmb$ is comparable in amplitude to the $\omb$ variations, with strong positive and negative oscillations in the derivative.
However, as already mentioned above, $\tcmb$ is not only already tightly constrained by \FIRAS \citep{Fixsen2009}, but a future CMB spectrometer will measure it directly from the CMB blackbody part, reaching the nK level precision in terms of statistical errors \citep{Abitbol2017}. This could possibly even allow us to measure the real-time change of the CMB temperature \citep{Abitbol2020}.
Thus, within standard cosmology, the CRR will not add any extra information.\footnote{Possible extensions would need to assume a change in the temperature-redshift relation or significant local gravitational effects to modify the observed value of $\tcmb$ from that expected by $T_\gamma=\tcmb(1+z)$ around recombination. However, this is hard to achieve without violating other distortions constraints, since thermalization is so inefficient \citep{Chluba2014TRR}.}

Cold dark matter density variations provide another unique imprint on the CRR. This is mainly through changes in the recombination dynamics, affecting both the time of recombination (shift in the line positions) and the duration (relative width of the recombination features). These general aspects are implicitly reflected in the shape of the log-derivative (Fig.~\ref{fig:parameters}). Broadly speaking, when $\omc$ is increased, the peaks of the CRR emission decrease whilst the minima see extra emission, leading to an overall smoothing of the recombination features.
The visible differences in comparison to the derivative with respect to $\omb$ are expected to make these two parameters in principle distinguishable.
For instance, there are notable changes in the oscillations of the two derivative spectra from the positive to negative branches, particularly in the trough between the H-$\alpha$ line and the Balmer continuum (Fig.~\ref{fig:parameters}). 
However, in comparison to the response from $\omb$, one expects the constraint to be $\simeq 10-20$ times weaker.

Since changes caused by variations of $\omc$ only impact the CRR through modifications to the expansion rate, $H(z)$, one could expect variations of $H_0$ to also lead to noticeable effects.
Indeed it has previously been shown that the response from $H_0$ is small but not totally negligible when $\Omega_{\rm b}h^2$ and $\Omega_{\rm tot}$ are kept fixed \citep{Chluba2008T0}. However, in this work we keep $\omc$ and $\omb$ fixed\footnote{Variations of $H_0$ are isolated by adjusting $\Omega_{\rm b}$ and $\Omega_{\rm cdm}$ such that $\omb$ and $\omc$ remain unchanged. We furthermore adjust $\Omega_\Lambda$ to keep a flat Universe, however, the CRR is hardly affected by the latter.}, such that the remaining constraining power on varying $H_0$ is found to be very weak and consistent with zero at the numerical precision of the computation [i.e., $\partial\dIv/\partial\ln\Ho\simeq (10^{-6}-10^{-5})\,\dIv$ see Fig.~\ref{fig:parameters}]. This implies that {\it no direct} constraint on $H_0$ can be expected from the CRR. This expectation is also supported by recent discussions of the recombination process in relation to the $H_0$ tension \citep{Ivanov2020}.

\begin{figure}
    \centering
    \includegraphics[width=\columnwidth]{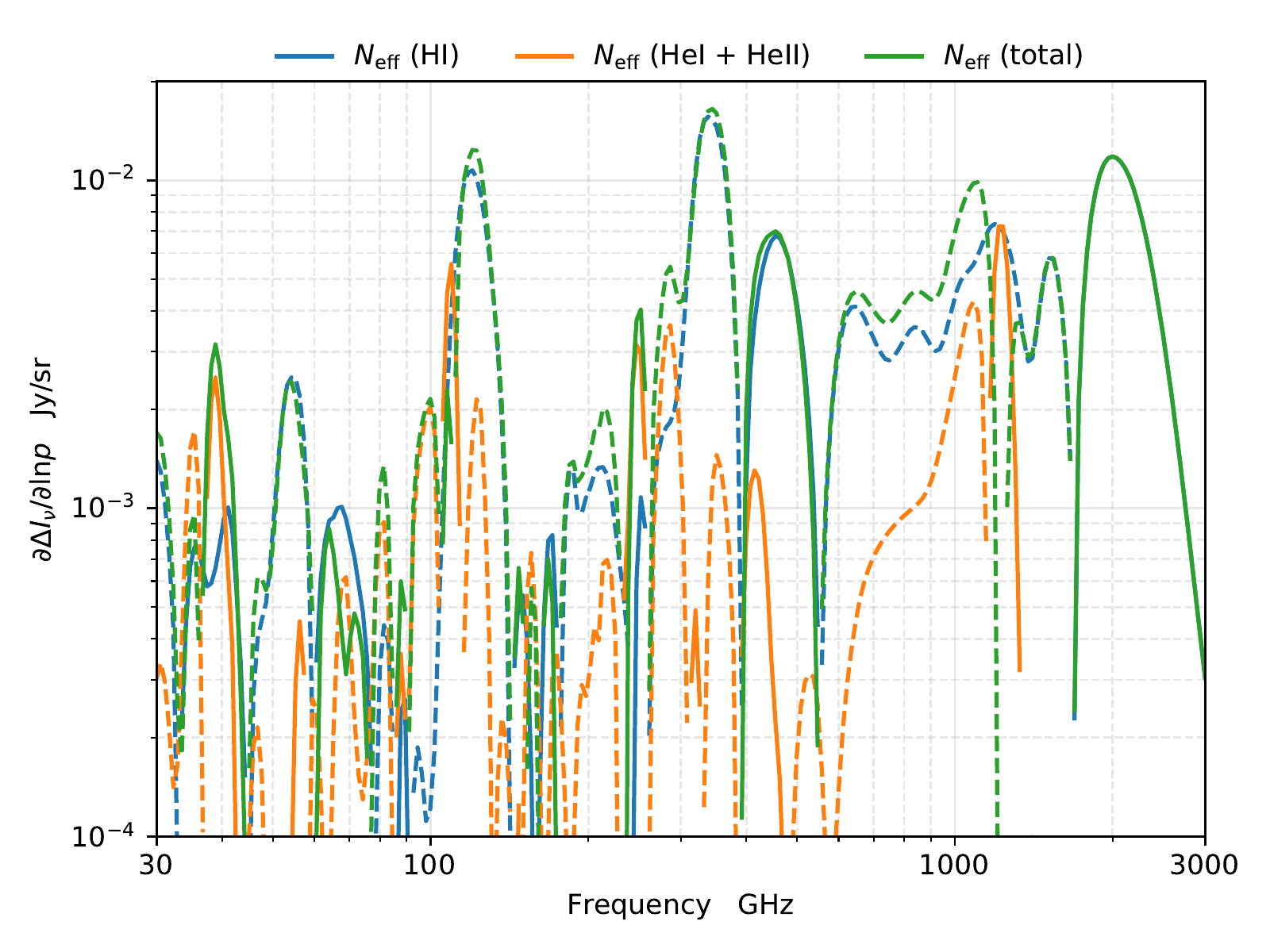}
    \vspace{-3mm}
    \caption{Separate contributions to the $\Neff$ response from hydrogen and the two helium recombination eras. The largest contribution comes from hydrogen, but locally helium can contribute significantly (e.g., around 1.1~THz).}
    \label{fig:Neff_direct}
    \vspace{-3mm}
\end{figure}

\vspace{-1mm}
\subsection{Extensions to $\Lambda$CDM}
\label{sec:extensions}
We can also study the effects of extensions to the $\Lambda$CDM model with the recombination radiation. One of the most interesting questions is how the abundance of primordial helium may differ at early times before recombination. This parameter can be directly constrained with the CMB anisotropies \citep{Planck2015params}, however, due to degeneracies with other parameters this proves challenging.
A more direct measurement can in principle be achieved using the CRR due to the intertwined nature of the hydrogen and helium emission lines. 
The $\Yp$ response of the CRR is shown in Fig.~\ref{fig:parameters}. Overall, the average signal is about one order of magnitude lower than the CRR, however, locally the response can be enhanced, e.g., around $\nu\simeq 250\,\GHz$. This amplification is caused by the dynamics of the neutral helium recombination lines and further enhanced by helium feedback.
Indeed, helium feedback processes cause a significant increase in the number of spectral features in the CRR, thereby making the response more easy to detect \citep[e.g.,][]{Chluba2009c, Chluba2012HeRec}. 

Like $\omc$, variations of $\Neff$ enter the recombination calculation through their effect on the expansion rate. Since during recombination, the medium is matter-dominated (this is not true for singly-ionized helium but its emission is diluted relative to hydrogen), the effect of $\Neff$ on the CRR is thus found to be sub-dominant \citep{Chluba2016CosmoSpec}. 
Comparing the relative contributions from hydrogen and the two helium recombination eras, we see that both are roughly equally important, each dominating in certain spectral bands (Fig.~\ref{fig:Neff_direct}). 
For example, at $\nu\simeq 1.1$~THz, the helium emission compensates the negative response from hydrogen and at 280~GHz, another helium feature is visible.
This implies that by distributing spectral measurements across frequency, one can in principle measure $\Neff$ at {\it independent times} during the recombination eras, but this task is clearly more futuristic.

We understand that variations in $\Neff$ affect the CMB anisotropies primarily through the expansion history \citep{Abazajian2015}. Together with the CRR, this may give us another tool to constrain the relativistic energy density. From the lower panel of Fig.~\ref{fig:parameters}, we can see that on average the $\Neff$ response is $\left|\partial\dIv/\partial\ln\Neff\right|\sim 0.1\times\left|\partial\dIv/\partial\ln\Yp\right|$. However, locally it can become comparable and also exhibits a different frequency pattern, making it in principle distinguishable. In addition, through the BBN relations (see Sect.~\ref{sec:BBN}), $\Neff$ can indirectly modify $\Yp$, thus increasing the related signal.
These types of analysis thus provide two independent constraints on $\Neff$, one focusing on BBN the other on recombination at $z\simeq 1100$, $2000$ and $6000$.

\vspace{-4mm}
\subsubsection{Impact of the 2s-1s two-photon decay rate}
\label{sec:extensions_A2s1s}
Variations in specific atomic transition rates have been studied using the CMB anisotropies to ascertain whether atomic physics during recombination can differ from the present day. One ideal candidate in this respect is the HI 2s-1s two-photon decay rate $\atwo$. Not only does it significantly control the dynamics of hydrogen recombination \citep{Zeldovich68}, it also has not been measured precisely in the lab \citep{Mukhanov2012}. Modern QED computations yield $\atwo=8.2206\, {\rm s}^{-1}$ \citep{Labzowsky2005} and thanks to precision CMB data from \Planck, we now have a cosmological measurement: $\atwo=\left(7.75\pm0.61\right)  {\rm s}^{-1}$ \citep{Planck2015params}. The variations in $\atwo$ are degenerate with $\ns$ and $\Ho$ for the CMB anisotropies. Since the CRR does not depend on $\ns$ and also only very weakly on $\Ho$, one can thus hope to improve the constraint on $\atwo$ using CMB spectroscopy.

The variations due to $\atwo$ are shown in the lower panel of Fig.~\ref{fig:parameters}. Overall, the level of the signal is comparable to that caused by changes in $\Yp$. The 2s-1s two-photon continuum distortion, which contributes direct emission around $\nu\simeq 1\,$~THz, depends explicitly on this parameter. Increasing the $\atwo$ parameter also accelerates the hydrogen recombination process, generating a unique response from all the hydrogen lines. Since for larger $\atwo$ more electrons can reach the ground-state of the hydrogen atom through the 2s-1s channel, this leads to a net reduction of transitions in the 2p-1s channel at high frequencies. This effect is visible as a negative feature around the Lyman-$\alpha$ line and also identifies parts of the CRR that on average are dominated by levels connecting through the 2p-1s channel to the ground state (regions of negative response). This thus provides a direct probe of the recombination dynamics.
Note that here we only vary $\atwo$ for hydrogen, and thus the helium lines are not affected by this. The importance of the two-photon decay channels is furthermore much lower for helium \citep{Chluba2009c}, such that a significantly reduced sensitivity to the corresponding helium rates is expected.

\subsubsection{Consistency relations from BBN}
\label{sec:BBN}
\begin{figure}
    \centering
    \includegraphics[width=\columnwidth]{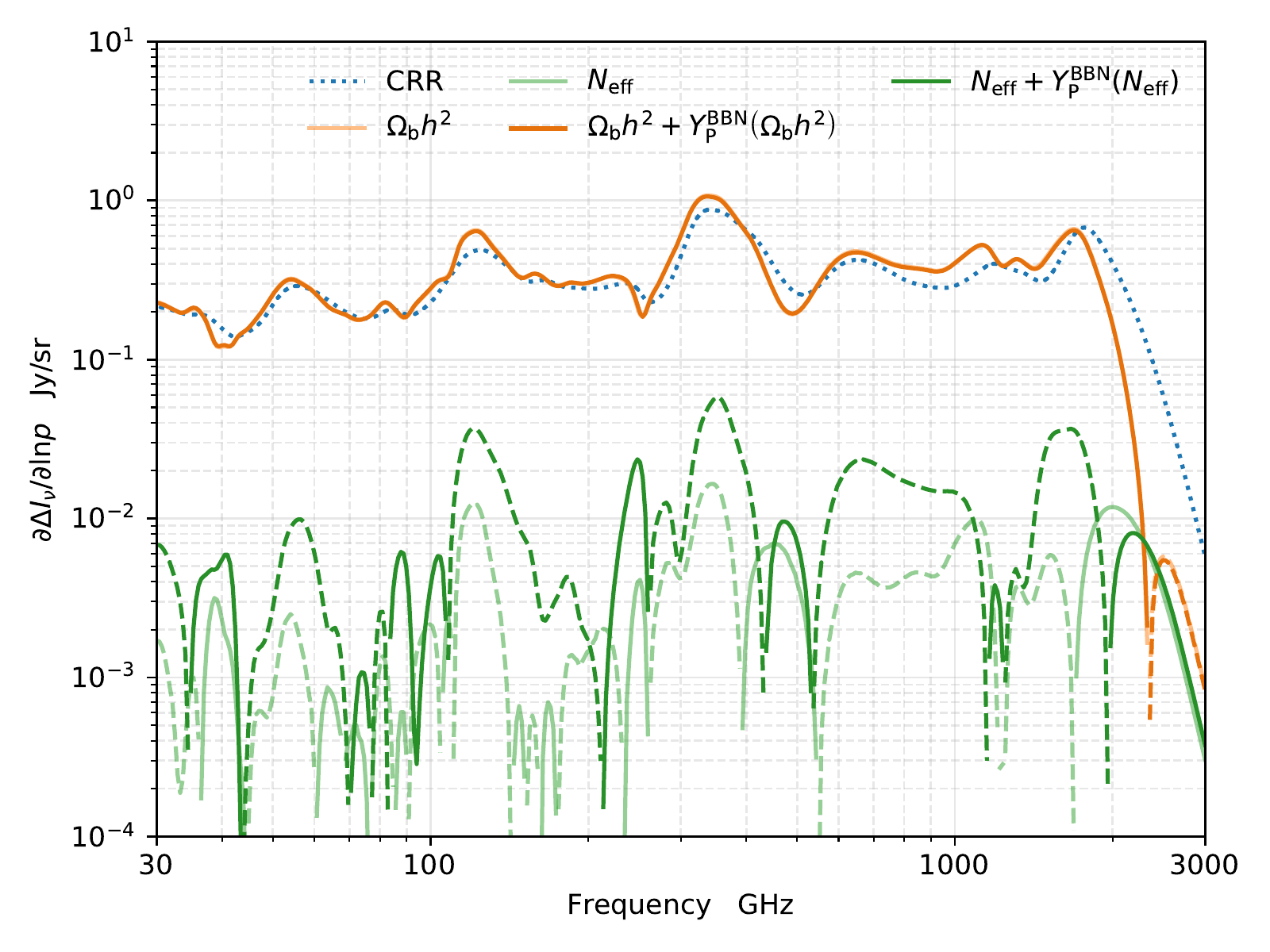}
    \caption{Comparison of the parameter responses of $\omb$ and $\Neff$ when we fully consider the consistency relation defined in Eq.~\eqref{eq:bbn} for primordial abundances during BBN. The faded curves match the standard parameter response curves shown in Fig.~\ref{fig:parameters}.}
    \label{fig:bbn}
\end{figure}

Rather than allowing $\Yp$ and $\Neff$ to freely vary, we can study how these parameters are related through BBN physics at $z\simeq 10^8$. In this, the baryon-photon ratio $\eta\equiv n_{\rm b}/n_{\gamma}\propto \omb/\tcmb^3$ plays a crucial role \citep{Steigman2007}. However, in this paper we will not explore the variations of $\tcmb$ given the precision of previous measurements \citep[e.g.,][]{Fixsen1996}. As a result, we follow similar methods used for the analysis of \Planck data and earlier works,  which utilizes the {\tt PArthENoPE} software \citep{PARTHENOPE} and interpolates over the results to find a relationship between $\Ypb$, $\omb$ and $\NeffBBN$ \citep{Iocco2009,Hou2013,Planck2015params}. Using these results, $\Ypb$ is given by:
\begin{equation}\label{eq:bbn}
    \Ypb =
    \begin{pmatrix}
    1 \\ \Delta\Neff\\ \Delta\Neff^2
    \end{pmatrix}^{\rm T}\,
    \begin{pmatrix}
    0.2311 &0.9502& -11.27 \\ 0.014& 0.009& -0.181 \\ -0.001& -0.001& 0.017
    \end{pmatrix}\,
    \begin{pmatrix}
    1 \\ \omega_{\rm b} \\ \omega_{\rm b}^2
    \end{pmatrix},
\end{equation}
where $\omega_{\rm b}\equiv\omb$ and $\Delta\Neff = \NeffBBN-3.046$. For small changes in $\omb$ and $\NeffBBN$, one can calculate the effective change in $\Yp$ due to changes in the background cosmology. In this way, the CRR parameter derivatives with respect to $\omb$ and $\Neff$ are modified. The results for these changes are shown in Fig.~\ref{fig:bbn}. 
The variations for the case with added helium abundance shifts are shown as $p_i+\Ypb (p_i)$ for $p_i=\{\omb, \NeffBBN\}$. 
Note that the respective changes without the added consistency relation for $\Ypb$ are shown as faded lines for each parameter. 
As expected, $\Ypb(\omb)$ has a small effect on the baryon density response. To leading order one can simply write
\begin{align}
\nonumber
\frac{\partial\dIvBBN}{\partial \ln\omega_{\rm b}}\approx \frac{\partial\dIv}{\partial \ln\omega_{\rm b}}+0.010\,\frac{\partial\dIv}{\partial \ln\Yp},
\end{align}
which explains why the effect is small (at the $\simeq 0.1-1\%$ level). The constraint on $\omega_{\rm b}$ is therefore not affected at a significant level.

However, the reverse can be said for changes in $\NeffBBN$, where the $\Ypb$ variation becomes the dominant source, e.g., at higher frequencies. 
Due to the smaller response from independently changing $\Neff$ in the background cosmology, compared to small changes in $\Yp$, this makes $\NeffBBN$ changes more noticeable:
\begin{align}
\nonumber
\frac{\partial\dIvBBN}{\partial \ln\Neff}\approx \frac{\partial\dIv}{\partial \ln\Neff}+0.043\,\frac{\partial\dIv}{\partial \ln\Yp}.
\end{align}
Given that $\partial\dIv/\partial \ln\Yp$ is about one order of magnitude larger than $\partial\dIv/\partial \ln\Neff$ (see Fig.~\ref{fig:parameters}), the BBN $\Neff$ response naturally borrows many spectral features of $\Yp$ (Fig.~\ref{fig:bbn}). Below, we present the results for various cases, showing that in particular when asking questions about extra relativistic species, this is important.

\vspace{-3mm}
\section{Setup of the analysis}
\label{sec:measurements}

\begin{figure}
    \centering
    \includegraphics[width=\columnwidth]{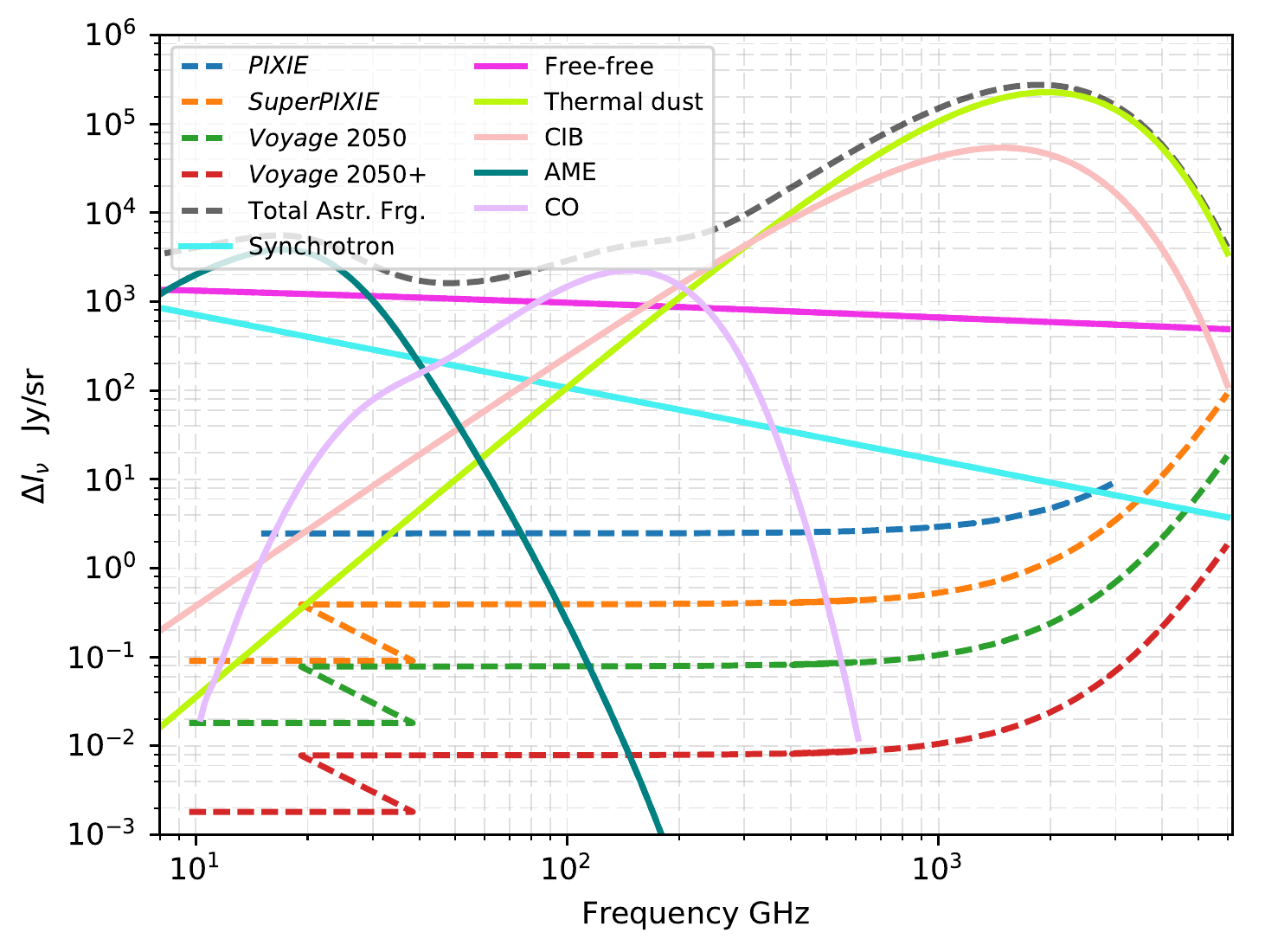}
    \includegraphics[width=\columnwidth]{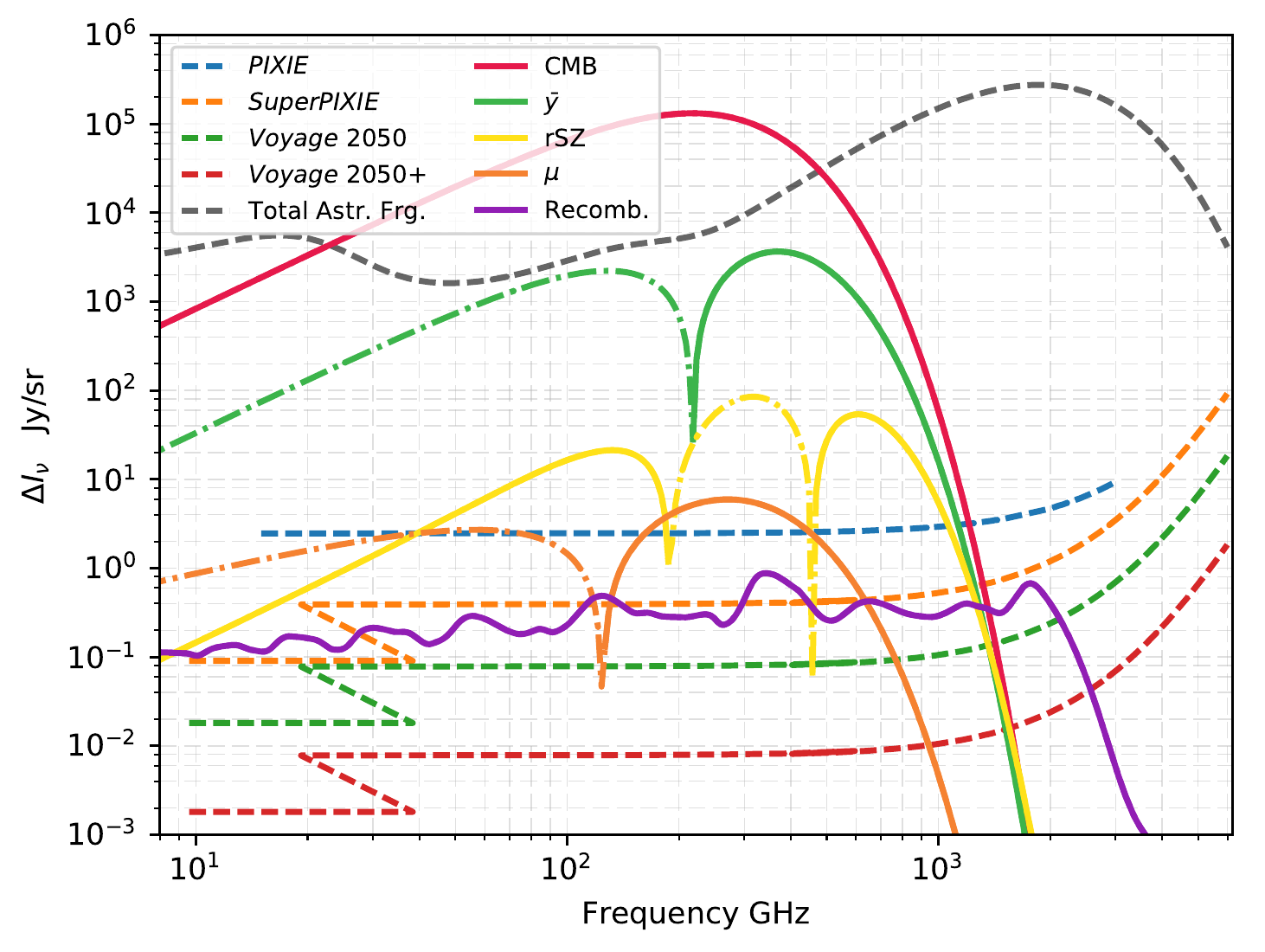}
    \caption{These figures depict the different spectral distortion signals, astrophysical foregrounds and the derivatives of the H and He recombination spectrum with respect to  cosmological parameters. Also depicted are the sensitivities of the proposed and futuristic spectrometers.}
    \label{fig:distortions}
\end{figure}

\subsection{Experimental concepts}
\label{sec:experiments}
%
Measurements of the CRR will be challenging, foremost owing to the low amplitude of the signal. In addition, foregrounds and even other expected CMB distortion signals heavily obscure the signal, necessitating wide frequency coverage, high sensitivity and extreme control of systematic effects.

Recent studies have shown that large improvements in sensitivities over \FIRAS can in principle be achieved using present-day technology. One proposal is \PIXIE \citep{Kogut2011PIXIE, Kogut2016SPIE}, which uses a highly symmetrized Fourier transform spectromter (FTS) approach derived from \FIRAS.
\PIXIE covers a frequency range of 15-3000~GHz with a bandwidth of 15 GHz per channel. For a 4-year mission, an all-sky sensitivity of $\simeq 2.5$~Jy/sr can be reached at low frequencies (see Fig.~\ref{fig:distortions} for details). 

Using multiple copies of \PIXIE, one can further improve the sensitivity, yielding \SuperPIXIE \citep{Kogut2019BAAS}.
It covers a more ambitious frequency range of 10-6000 GHz through a modular design with three different FTS at low (10-40 GHz, $\Delta \nu$=2.4~GHz), mid (20-600~GHz, $\Delta \nu =19.2$~GHz) and high (400-6000 GHz, $\Delta I_{\nu}=57.6$ GHz) frequencies. The sensitivities depicted in Fig.~\ref{fig:distortions} are derived assuming the default 4-4-1 configuration (i.e, 4 copies of both low and mid frequency modules and 1 copy of the high frequency module) over an 8-year mission. This configuration was chosen for mitigating low-frequency foreground challenges hampering a detecting of the $\mu$ distortions \citep[see][for details]{Abitbol2017, Kogut2019BAAS}. 

\Voyage is a spectrometer concept derived from \SuperPIXIE with identical specifications, except an overall enhancement in sensitivity by a factor of 5. This can in principle be achieved by further multiplying \PIXIE, increasing the sizes of the individual telescopes and detectors, or extending the observing time \citep{Chluba2019BAAS}.
Anticipating major improvements of experimental and space-exploration capabilities \citep[e.g.,][]{Silk2018Nature}, here we study even more futuristic constraints beyond \Voyage.
We find that foreground mitigation for the CRR differs from what is required for precise measurements of $\mu$ and $y$. We thus expect that additional gains could be made by optimizing the frequency range and other experimental parameters, although a detailed exploration is beyond the scope of this work.
Taking this into account, for \Voyagep we optimistically assume another 10 times higher sensitivity with respect to \Voyage. 
As we show here, \Voyagep will already enable deriving competitive cosmological parameter constraints from measurements of the CRR.

\vspace{-3mm}
\subsubsection{Bandwidth averaging}
\label{sec:bandwidth}
The constraints are a consequence of subtle shifts in the features of the CRR due to varying cosmological parameters. These features can only be robustly measured when accounting for the presence of instrument band averaging, which is typically smaller than the size of the features in the spectrum. The width of the features in the CRR changes with frequency (Fig.~\ref{fig:parameters}) and it is not easy to directly infer the frequency range responsible for most of the constraining power. When deriving the forecasts we take care to duly account for the band averaging with specifications provided in Sect.~\ref{sec:experiments}. However, for the given specification, we find no significant effect on the forecasts. 

\vspace{-3mm}
\subsubsection{Systematic effects}
\label{sec:systematics}
Aside from the aforementioned instrumental effects and various foregrounds (next section), measurements of the CRR (and spectral distortions more generally) will be strongly affected by control of systematics. One of the most important systematics is the calibration of the channels, another comes from the beams and their stability. With spectrometer concepts like \PIXIE, a high level of build-in symmetries allow controlling these to unprecedented accuracy \citep{Kogut2016SPIE, Kogut2018beams, Kogut2019System, Kogut2020Cali}.
While for concepts like \SuperPIXIE and \Voyage, it is plausible that the required control of systematics can be reached with present-day technology, going beyond will likely need more work. 
Here, it is important to mention that due to the significant frequency-dependence of the CRR, a frequency-to-frequency comparison may allow extracting this signal, thus not necessarily requiring absolute calibration \citep{Sunyaev2009, Mukherjee2019}. On the other hand, accurate channel-intercalibration will likely demand a precise reference load, essentially making this equivalent to absolute calibration.
Given the approximate nature of our study, we shall assume that systematics are fully under control, deferring a detailed treatment to future work.

\subsection{Astrophysical foregrounds}
\label{sec:foregrounds}
In real world setting, one invariably has to model and subtract the galactic and extra-galactic foregrounds to make measurements of the CRR possible. For our Fisher estimates, we include the main astrophysical foregrounds relevant to measurements of the CMB monopole spectrum. We follow the method of \citet{Abitbol2017}, who carefully studied the importance of foregrounds to measurements of expected $\mu$- and $y$-type distortions. Their model includes commonly used parametrizations for low- (synchrotron\footnote{Note that we do not include the running of synchrotron spectral index in this work, as this has more bearing to the $\mu$ distortion measurement forecasts.}, AME, free-free, integrated CO) and high-frequency foregrounds (thermal dust and CIB). We refer to \citet{Abitbol2017} for details, but in total this yields 11 parameters to marginalize over. We will refer to these as `astrophysical foreground' parameters.

We immediately emphasize that at the level of precision required for a detection of the CRR, we have {\it no idea} if these simple parametrizations will be applicable. Thus, our estimates remain optimistic limits and should be interpreted as such.
Extended parametrizations of foregrounds to include the inevitable and expected effects of various averaging processes can in principle be achieved using moment expansion methods \citep{Chluba2017}. To subtract monopole foregrounds using CMB maps, a recent study approached this problem by combining moment with classical ILC methods \citep{Rotti2020}; however, we omit these effects here. 
As we will show below, unlike the broad $\mu$ and $y$-type distortions, the observability of the CRR is less hampered by foregrounds due to the unique frequency-dependence of the recombination features. This builds additional confidence with regards to the detectability of the CRR using future spectrometers, and could also allow us to simplify the experimental approach. For example, observations from the ground it was shown that measurements over one octave in frequency could still yield a detection of the CRR \citep{Mayuri2015}.

We also highlight that the role of foreground priors is marginal when considering the CRR. For detections of primordial $\mu$ distortion, coverage at frequencies below the \PIXIE range was shown to be very important \citep{Abitbol2017}. From space this is difficult to acheive due to constraints on the size of the detector. However, ground-based observations could potentially help in this respect, enabling better control of the synchrotron and free-free foregrounds for the extraction of $\mu$. We find that for the CRR, this does not seem to be required, providing another opportunity for optimization.

\subsection{Other CMB spectral distortion signals}
\label{sec:other_SD}
In addition to the astrophysical foregrounds mentioned above, to obtain reliable limits on the CRR we also need to include the other expected CMB distortion signals (see Fig.~\ref{fig:distortions}). An overview of all the expected signals can be found in \citet{Chluba2016}. Additional broad discussions were presented in \citet{Chluba2011therm}, \citet{Sunyaev2013} and \citet{Lucca2020}.

The largest distortion is due to the cumulative effect of all SZ clusters in the Universe \citep{Refregier2000, Hill2015}. This yields a $y$-type distortion with average $y\simeq \pot{1.8}{-6}$ using most recent parameters for the halo model \citep{Hill2015}. Since the typical temperature of the clusters contributing to this signal is $k\Te \simeq 1\keV-3\keV$, relativistic temperature corrections (referred to as rSZ) become noticeable \citep{Hill2015}. The precise value of the average electron temperature depends on the details of the temperature-mass scaling relations \citep[e.g.,][]{Remazeilles2019, Lee2020}, but here we will assume $k\Te =1.3\keV$, following \citet{Hill2015}. Together these signals probe the hot gas in the Universe \citep[see][for more discussion]{Chluba2019BAAS}.

The last inevitable distortion is caused by the damping of small-scale acoustic waves in the pre-recombination era \citep{Sunyaev1970diss, Daly1991, Hu1994}. This signal can now be computed accurately \citep{Chluba2012, Pajer2012b, Inogamov2015} yielding $\mu\simeq \pot{2.3}{-8}$ for standard slow-roll inflation. The exact signal also has contributions from $y$ and furthermore is modified by the smaller baryonic cooling distortion \citep{Chluba2005, Chluba2011therm}, but these can be neglected for our purposes here. This signal can be used to constrain inflation models \citep{Chluba2012inflaton, Khatri2013forecast, Chluba2013fore, Chluba2013PCA, Cabass2016}.

Finally, in addition to the CMB distortion parameters, $y, \mu$ and $k\Te$, one also has to vary the value if the CMB monopole temperature, $\tcmb$. Its precise value is not known to the level of precision expected from future CMB spectrometers (down to nK), and the constraint will be directly driven by the blackbody part of the CMB spectrum. Together this generally means that a total of four CMB spectral parameters have to be added to the forecast. We will refer to them as `distortion foregrounds'. 

Future SZ cluster measurements and more detailed theoretical modeling of the SZ contribution may allow further constraining the average $y$-parameter. In addition, we can expect the relativistic SZ contribution to be at least theoretically modeled better. And finally, assuming standard slow-roll inflation we can in principle precisely model the contributions from $\mu$. Together, this could potentially improve the expected detection limits when studying the CRR. To bracket the range we will thus present forecast for various spectral and foreground parameter combinations.

\subsection{Simple recap of the Fisher method}
\label{sec:fisher_method}
We model the total spectrum measured on the full sky as follows,
\begin{equation}
\Delta I_{\nu}^{\rm Total} = \Delta I^{\,\rm rec}_{\nu} + \sum_f^{\rm Sync,\, Dust \cdots}\Delta I^f_{\nu} + \sum_s^{y, \,\mu \cdots}\Delta I^s_{\nu} \,.
\end{equation}
With this data model, we assume that only $\Delta I^{\,\rm rec}_{\nu}$ has a cosmological parameter dependence, while the 15 parameters characterizing all other components of the spectrum are treated as nuisance parameters.
Our primary aim is to forecast the constraint on cosmological parameters from measurements of CRR.
While the spectral distortions signals ($\mu$, $y$, $k\Te$) also have some cosmological parameter dependence, these spectra are smooth as compared to CRR (i.e they  have fewer features), making them more susceptible to astrophysical foregrounds. Additionally, the changes in the spectral shapes due to variation in cosmological parameters are more subtle and consequently their ability to constrain these parameters is expected to be weak. 
This does not apply to $\mu$-distortion constraints on the primordial power spectrum, however, the interplay is indirect and only important once concepts beyond \Voyagep are considered.
We therefore proceed assuming that with this spectral measurement, the dominant gains in cosmological parameters constraints will arise from measurements of CRR. 

Under the assumptions of a Gaussian likelihood, the Fisher matrix encodes complete information on the parameter errors and their covariances and can therefore be used to derive parameter constraints. In our setting, the Fisher matrix is estimated as follows,
\begin{equation}
\mathcal{F}_{pp'} = \sum _{\nu \nu'}\frac{\partial \Delta I^{\rm Total}_{\nu}}{\partial p} \mathcal{C}^{-1}_{\nu \nu'}  \frac{\partial \Delta I^{\rm Total}_{\nu'}}{\partial p'} \,,
\end{equation}
where $\mathcal{C}_{\nu \nu'}$ denotes the channel covariance matrix, which describes the measurement sensitivity (squared) and is assumed to be diagonal (i.e., measurements across channels are independent). 

We construct the Fisher matrix, by first injecting only the CRR spectrum to assess the statistical significance of merely its detection. For experimental setting where the CRR is detected at high significance we asses the cosmological parameter constraint. For this case, we evaluate the Fisher matrix with derivatives of the CRR with respect to the cosmological parameters as input, details of which are summarized in Sec.~\ref{sec:standard}. Maintaining consistency with analysis in \cite{Chluba2019Voyage}, we use a 1\% prior of parameters characterizing the low frequency foregrounds (i.e synchrotron and free-free), however we find these have no bearing on the forecasts derived for the CRR.

In addition to providing the spectrometer only forecasts we also present results from a joint forecasting analysis by explicitly incorporating the full \Planck parameter covariances. These \Planck priors were sourced from the marginalised covariance matrices of MCMC runs using {\tt CosmoMC} with \Planck 2018 data\footnote{We use the \Planck 2018 TTTEEE+low$\ell$+lowE+lensing data.}
\citep{COSMOMC,Planck2018like}. Specifically, these constraints come from running $\omb$, $\omc$, $\thetaMC$, $\tau$, $\ns$ and $\As$ with extension parameters $\Yp$, $\Neff$ and $\atwo$ as described in Sect.~\ref{sec:forecasts}.
The forecasts are repeated for a varying number of nuisance parameters to precisely understand the degradation on the CRR forecasts resulting from other CMB signals, astrophysical foregrounds and finally from combining all parameters.

\vspace{2mm}
\section{Cosmology from measurements of the CRR}
\label{sec:forecasts}
In this section, we present our CRR forecasts. We first discuss the detectability, using the CRR as a fixed spectral template. While \PIXIE is found to be short of sensitivity, \SuperPIXIE, \Voyage and \Voyagep allow in principle measurements of the signal and hence are studied further.
We present cosmological parameter constraints for spectrometer-only measurements as well as in combination with existing CMB anisotropy data.

\begin{table}
\centering
\begin{tabular}{lcccc}
\toprule
Analysis & \PIXIE & \SuperPIXIE & \Voyage & \Voyagep    \\
\midrule
No Frgs. & 1.6 & 9.5 & 48 & 477 \\
Dist. Frgs. & 1.1 & 3.6 & 18 & 179 \\
Astr. Frgs. & 0.5 & 2.5 & 12 & 122 \\
$\textbf{All Frgs.}$ & 0.3 & 1.5 & 7.7 & 77 \\
\bottomrule
\end{tabular}
    
\caption{Detection of the amplitude of the recombination spectrum for various spectrometers for different assumptions about foregrounds. We give the signal-to-noise ratio for the CRR amplitude.}
\label{tab:detect}
\end{table}
\subsection{Detectability of the CRR}
\label{sec:detection}
The simplest question to ask is how well a given experiment can detect the overall amplitude of the CRR. The results of this exercise are summarized in Table~\ref{tab:detect}. Even in the absence of any foregrounds, \PIXIE itself only achieves a marginal limit on the amplitude of the CRR. While the detailed numbers depend on the total survey time\footnote{Here, we assumed 4 years but factors $\simeq \sqrt{2}$ can be gained with an extended mission \citep{Kogut2016SPIE}.}, the limit further degrades when adding foregrounds, ultimately yielding an SNR of $\simeq 0.3$. For this reason, we will not consider \PIXIE itself any further in our forecast.

At the sensitivity envisioned for \SuperPIXIE, a nearly $10\sigma$ detection of the CRR could be expected in the absence of foregrounds.
However, once foregrounds are added, the constraint weakens, dropping to $\simeq 1.5\sigma$ when all foreground parameters are varied. 
It is thus clear that even \SuperPIXIE will not allow to constrain cosmological parameters using the CRR to high significance. For comparison, we will quote a few numbers in the more extensive forecasts below, but generally we also omit \SuperPIXIE in the main analysis.

For \Voyage and \Voyagep, large additional factors for the SNR can be gained due to their enhanced sensitivity. Even when adding all foregrounds, \Voyage and \Voyagep still achieve significant detections of the CRR. 
The degradation by foregrounds seems to be driven to similar parts by distortion foregrounds (four parameters) and astrophysical foregrounds (11 parameter). This highlights that the CRR benefits from many spectral features that can help identifying it. 
The final estimates for \Voyage agree well with what was presented in \citet{Chluba2019BAAS}, and large gains in constraining power (reaching a $\simeq 80 \sigma$ detection of the CRR) can be expected for \Voyagep.

\begin{table}
\centering
\begin{tabular}{llcc}
\toprule
Expt. & Analysis &  $\omb$ &  $\omc$ \\
\midrule
\multirow{3}{*}{\Voyage} & No Frgs. &            44 &                     2.1  \\
            & Dist. Frgs. &            14 &                     1.3  \\
            & Astr. Frgs. &             7.4 &                     0.9  \\
            & $\textbf{All Frgs.}$ &             4.8 &                     0.7  \\
\cline{1-4}
\multirow{3}{*}{\Voyagep} & No Frgs. &           445 &                    21  \\
            & Dist. Frgs. &           137 &                    14 \\
            & Astr. Frgs. &            74 &                     8.9  \\
            & $\textbf{All Frgs.}$ &            48 &                     6.7  \\
\bottomrule
\end{tabular}
    
\caption{Spectrometer only constraint on $\omb$ and $\omc$. We quote the signal-to-noise ratio (SNR) with respect to the fiducial value.}
\label{tab:Spec_case_I}
\end{table}

\begin{table}
\centering
\begin{tabular}{llccc}
\toprule
Expt. & Analysis &  $\omb$ &  $\omc$ &  $\Yp$   \\
\midrule
\multirow{4}{*}{\Voyage} & No Frgs. &            38 &                     1.8 &      4.8 \\
            & Dist. Frgs. &            12 &                     1.3 &     3.9 \\
            & Astr. Frgs. &             6.3 &                     0.9 &       2.8 \\
            & $\textbf{All Frgs.}$ &             4.8 &                     0.6 &      2.3 \\
\cline{1-5}
\multirow{4}{*}{\Voyagep} & No Frgs. &           381 &                    18 &     48 \\
            & Dist. Frgs. &           116 &                    13 &      39 \\
            & Astr. Frgs. &            63 &                     8.6 &       28 \\
            & $\textbf{All Frgs.}$ &            48 &                     5.9 &       23 \\
\bottomrule
\end{tabular}
    
\caption{Like Table~\ref{tab:Spec_case_I} but for $\omb$, $\omc$ and $\Yp$.}
\label{tab:Spec_case_II}
\end{table}

\subsection{Spectrometer-only constraints from the CRR}
\label{sec:fisher_standard}
We now present the estimated constraints independently obtained by measuring the CRR with \Voyage and \Voyagep. For the first case, we assume that $\Yp$ and $\Neff$ are fixed to their standard values. 
The constraints on $\omb$ are not expected to be affected by this approximation, as the extra correction using $\Ypb(\omb)$ will not modify the CRR response with respect to $\omb$ much (Fig.~\ref{fig:bbn}). 
The results are summarized in Table~\ref{tab:Spec_case_I}. No explicit constraint on $H_0$ can be obtained, and thus was not included here. 
Impressive constraints on $\omb$ and $\omc$ can be obtained in the absence of foregrounds.\footnote{For \SuperPIXIE, we find that $\omb$ can be measured to $\simeq 10\%$ precision without foregrounds, but with foregrounds the SNR drops below unity.}
For $\omb$, these limits weaken by about one order of magnitude once all foregrounds are added. This is because the main effect of $\omb$ is a change in the amplitude of the CRR (Fig.~\ref{fig:parameters}), such that the overall sensitivity is similar to that of pure template fitting. Still, an independent $\simeq 5\sigma$ measurements of $\omb$ can be achieved with \Voyage alone. The constraint from \Voyagep is essentially one order of magnitude tighter, yielding a very precise measurement just from the CRR alone. In comparison to the CMB constraint from \Planck 2018, the CRR limit from \Voyagep is only a factor of $\simeq 2-3$ weaker, even in the presence of foreground.
Consistent with the reduced response to changes of $\omc$ (Fig.~\ref{fig:parameters}), the related SNR is another $\simeq 10-20$ times weaker than for $\omb$. Thus \Voyagep only warrants an independent constraint at the level of a few standard deviations once foregrounds are added. Overall, to become competitive to existing CMB constraints (errors at the level of $\simeq 1\%$) on $\omb$ and $\omc$ additional enhancements in sensitivity over \Voyagep are thus required.

\begin{table}
\centering
\begin{tabular}{llccc}
\toprule
Expt. & Analysis &  $\omb$ &  $\omc$ &  $\NeffBBN$             \\
\midrule
\multirow{4}{*}{\Voyage} & No Frgs. &            39 &                     1.4 &               0.8 \\
            & Dist. Frgs. &            12 &                     1.2 &               0.7 \\
            & Astr. Frgs. &             6.3 &                     0.7 &            0.5 \\
            & $\textbf{All Frgs.}$ &             4.8 &                     0.5 &               0.4 \\
\cline{1-5}
\multirow{4}{*}{\Voyagep} & No Frgs. &           389 &                     14 &              8.3 \\
            & Dist. Frgs. &           117 &                     12 &       6.8 \\
            & Astr. Frgs. &            63 &                     7.4 &               4.6 \\
            & $\textbf{All Frgs.}$ &            48 &                     5.0 &            3.8 \\
\bottomrule
\end{tabular}
    
\caption{Like Table~\ref{tab:Spec_case_I} but for $\omb$, $\omc$ and $\NeffBBN$. The BBN consistency relation is used to set $\Yp\equiv \Ypb(\omb, \NeffBBN)$.}
\label{tab:Spec_case_III}
\end{table}

\begin{table}
\centering
\begin{tabular}{llcccc}
\toprule
Expt. & Analysis &  $\omb$ &  $\omc$ &  $\Yp$ &  $\Neff$           \\
\midrule
\multirow{4}{*}{\Voyage} & No Frgs. &            36 &                     0.6 &      4.8 &            0.1 \\
            & Dist. Frgs. &            12 &                     0.5 &        3.9 &            0.1 \\
            & Astr. Frgs. &             6.3 &                     0.4 &      2.5 &            0.1 \\
            & $\textbf{All Frgs.}$ &             4.8 &                     0.3 &        2.0 &            0.1 \\
\cline{1-6}
\multirow{4}{*}{\Voyagep} & No Frgs. &           361 &                     5.5 &       48 &            1.2 \\
            & Dist. Frgs. &           116 &                     5.0 &     39 &            1.1 \\
            & Astr. Frgs. &            63 &                     3.9 &       25 &            1.0 \\
            & $\textbf{All Frgs.}$ &            48 &                     3.3 &       20 &            1.0 \\
\bottomrule
\end{tabular}
    
\caption{Like Table~\ref{tab:Spec_case_I} but for $\omb$, $\omc$, $\Yp$ and $\Neff$.}
\label{tab:Spec_case_IV}
\end{table}

In Table~\ref{tab:Spec_case_II}, we now also vary $\Yp$ as an independent parameter. The overall picture remains the same and only a small degradation of the constraint on $\omb$ and $\omc$ with respect to the previous case is found. This highlights the 'orthogonality' of the $\Yp$ response with respect to the dependence of the CRR on the other parameters. With \Voyagep, an impressive $\simeq 4\%$ measurements of $\Yp$ could be achieved even in the presence of all foregrounds. The degradation is only a factor of $\simeq 2$ in comparison to the case without foregrounds, again highlighting the unique spectral signal related to variations of $\Yp$, with many positive and negative features identifying it (Fig.~\ref{fig:parameters}).
To put these values into perspective, the $\Yp$ constraint from \Planck 2018 yields an error at the level of $\simeq 5.5\%$. 
    We can thus naively expects a factor of $\simeq 1.7$ improvement when adding a future spectrometer like \Voyagep to existing CMB data. Indeed, we find that noticeable improvements on $\Yp$ are the first to be seen in comparison to current constraints. As we will show below, the exact number is slightly higher, since $\omb$ also adds some extra information, helping to break degeneracies.

We next consider constraints on $\Neff$ using the CRR. Simply adding it as an independent parameter to the estimation problem yields extremely weak constraints even for \Voyagep (no more than $\simeq 1\sigma$). This is not surprising given the low amplitude of the corresponding CRR response (Fig.~\ref{fig:parameters}).
However, we can ask the question how well the presence of extra relativistic degrees for freedom at BBN can be constrained by propagating the effect to $\Yp$. The results of this exercise are presented in Table~\ref{tab:Spec_case_III}, where we used the BBN consistency relation (Sect.~\ref{sec:BBN}).
In this case, \Voyagep can provide an independent constraint at the level of $\simeq 4\sigma$ using a spectrometer alone. While not competitive with existing CMB constraints, it is clear that with additional improvements in the spectrometer sensitivity by a factor of $\gtrsim 4$ over \Voyagep one could in principle supersede the corresponding \Planck 2018 constraint.

\begin{figure}
    \centering
    \includegraphics[width=\columnwidth]{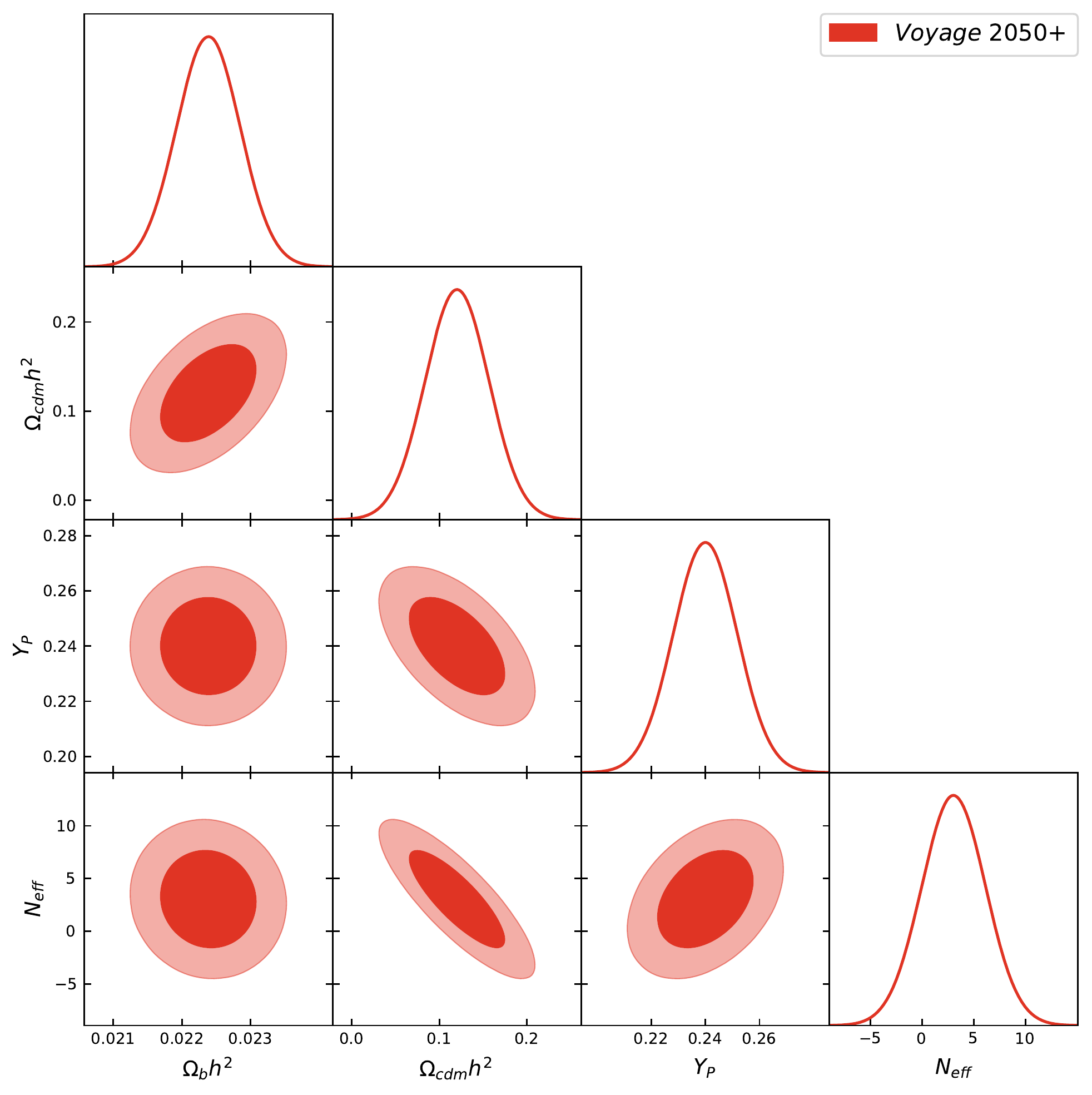}
    \caption{Spectrometer only parameter contours for \Voyagep.}
    \label{fig:contour_v2050X10_spec}
\end{figure}

As another example we vary all parameters $\omb$, $\omc$, $\Yp$ and $\Neff$ simultaneously (Table~\ref{tab:Spec_case_IV}). The corresponding parameter contours are shown in Fig.~\ref{fig:contour_v2050X10_spec} to illustrate the covariances. As expected, no competitive constraint on $\Neff$ can be obtained even for $\Voyagep$; however, the degradation of the errors on the other parameters is  marginal, so that together with CMB anisotropy data one could still expect gains on $\Neff$ due to the extra constraining power on $\Yp$. We indeed find a marginal improvement below. 

\vspace{2mm}
From Fig.~\ref{fig:contour_v2050X10_spec}, we can see that for CRR-only constraints, $\Yp$ does not correlate with $\omb$ and only slightly with $\omc$ and $\Neff$. Furthermore, $\omb$ only correlates with $\omc$.
This means that just adding information on $\omc$ would be sufficient to improve the limits on $\omb$ from \Voyage. Conversely, \Voyagep can help break some of the degeneracies and thus feed back information to the CMB anisotropy constraints, as demonstrated below. 
We note that using the \Voyagep results, one can easily scale the parameter errors with sensitivity to determine when interesting thresholds in comparison to existing or planned CMB anisotropy experiments can be expected.
Also, given that the errors on $\Neff$ and $\omc$ are still very large for measurements with \Voyagep alone, it is clear that the posteriors shown in Fig.~\ref{fig:contour_v2050X10_spec} are expected to exhibit non-Gaussian features. However, these are not captured with our Fisher method, as the modeling of the CRR no longer is truly perturbative.

\begin{table}
\centering
\begin{tabular}{llcccc}
\toprule
Expt. & Analysis  &  $\omb$ &  $\omc$ &  $\atwo$ \\
\midrule
\multirow{4}{*}{\Voyage} & No Frgs. &            45 &                     0.4 &         1.3 \\
                            & Dist. Frgs. &            14 &                     0.4 &         1.2 \\
                            & Astr. Frgs. &             7.2 &                     0.3 &         1.2 \\
                            & {\bf All Frgs.} &             4.8 &                     0.3 &         1.1 \\
\cline{1-5}
\multirow{4}{*}{\Voyagep} & No Frgs. &           445 &                     3.9 &        13 \\
                            & Dist. Frgs. &           137 &                     3.7 &        12 \\
                            & Astr. Frgs. &            72 &                     3.2 &        12 \\
                            & {\bf All Frgs.} &            48 &                     3.0 &        11 \\
\bottomrule
\end{tabular}
\caption{Like Table~\ref{tab:Spec_case_I} but for $\omb$, $\omc$ and $\atwo$.}
\label{tab:Spec_case_V}
\end{table}

Finally, to illustrate the direct constraining power of the CRR on recombination physics, we also considered a simple one parameter extension using $\atwo$. The corresponding spectrometer only result is presented in Table~\ref{tab:Spec_case_V}. \Voyagep would allow a $\simeq 10\%$ measurement of $\atwo$. Improving the sensitivity by another factor $\simeq 1.5$ would allow one to match the constraint obtained from \Planck, which give a $\simeq 6.6\%$ measurement \citep{Planck2015params}. This could in principle shed independent light on the dynamics of hydrogen recombination, highlighting how the CRR indeed is a probe of physics before last scattering \citep{Chluba2008T0, Sunyaev2009}.

\subsection{Combination with CMB anisotropy measurements}
\label{sec:fisher_plus_CMB}
From the results of the previous section we can already understand that even a spectrometer similar to \Voyage will be unable to improve the cosmological parameter constraints in comparison to \Planck. We checked this carefully by feeding current best \Planck constraints and covariances into our Fisher approach, finding negligible improvements in all cases (the results were all dominated by the priors from \Planck and even neglecting foregrounds did not affect the conclusion). 
However, for even more futuristic spectrometer concepts like \Voyagep one can expect interesting gains, driven by improvements on $\omb$ and $\Yp$ from the CRR.
The results of all our runs are summarized in Table~\ref{tab:Spec_Planck}. For cases without foregrounds, indeed a combination of the spectrometer with \Planck could yield interesting improvements for all the considered parameters, in particular for $\omb$ and $\Yp$, reaching $\Delta\omb/\omb \simeq 0.2\%$ and $\Delta\Yp/\Yp \simeq 1.7\%$. In these cases, this would even be on par with the improvements expected from future CMB experiments like the Simons Observatory \citep[SO,][]{SOWP2018}, which expect to achieve $\Delta\omb/\omb \simeq 0.2\%$ and $\Delta\Yp/\Yp \simeq 3\%$ \citep{SOWP2018}.

\begin{table}
\centering
\begin{tabular}{lccccc}
\toprule
Analysis &  $\omb$ &  $\omc$ &  $\Ho$ &  $\Yp$ &  $\Neff$  \\
\midrule
No Frgs. &            0.27 &                    0.78 &   0.66 & -- & --\\
Dist. Frgs. &            0.52 &                    0.80 &   0.72 & -- & --\\
Astr. Frgs. &            0.62 &                    0.83 &   0.76& -- & -- \\
$\textbf{All Frgs.}$ &            0.77 &                    0.88 &   0.84& -- & -- \\
\cline{1-6}
No Frgs. &            0.28 &                    0.82 &   0.62 &   0.31 & --\\
Dist. Frgs. &            0.60 &                    0.86 &   0.74 &   0.40 & --\\
Astr. Frgs. &            0.77 &                    0.92 &   0.86 &   0.45 & --\\
 $\textbf{All Frgs.}$ &            0.78 &                    0.93 &   0.86 &   0.55 & --\\
\cline{1-6}
No Frgs. &            0.24 &                    0.67 &   0.51 &     -- &      0.56 \\
Dist. Frgs. &            0.60 &                    0.83 &   0.76 &    -- &       0.80 \\
Astr. Frgs. &            0.80 &                    0.89 &   0.89 &      -- &    0.90 \\
$\textbf{All Frgs.}$ &            0.86 &                    0.92 &   0.91 &      -- &     0.92 \\
\cline{1-6}
No Frgs. &            0.26 &                    0.58 &   0.46 &   0.26 &           0.49 \\
Dist. Frgs. &            0.63 &                    0.65 &   0.62 &   0.30 &           0.59 \\
Astr. Frgs. &            0.83 &                    0.72 &   0.72 &   0.35 &           0.67 \\
$\textbf{All Frgs.}$ &            0.86 &                    0.79 &   0.80 &   0.45 &           0.75 \\
\bottomrule
\end{tabular}
    
\caption{Cosmological constraints from a combination of \Voyagep with \Planck. The numbers are quoted as ratio of the obtained errors relative to the \Planck-only error. If \Planck priors dominate, we obtain a ratio $\simeq 1$, and improvements are found when the quoted value is $<1$, with an improvement factor given by the inverse of the value. For the case with $\Neff$ added alone, we used the BBN consistently relation to set the value of $\Yp$.}
\label{tab:Spec_Planck}
\end{table}

However, once foregrounds are added the most interesting parameter is $\Yp$, for which an improvements by a factor $\simeq 2$ can be expected, still yielding $\Delta\Yp/\Yp \simeq 3\%$. When also adding $\Neff$ as independent parameter, the conclusion remains similar and  marginal improvements on the other parameters are also found, thus highlighting how a combination of \Planck with \Voyagep can obtain promising constraints that are competitive with those from planned CMB experiments. 
Most importantly, the degradation by foregrounds seems to be quite limited. This makes measurements of the CRR an interesting avenue for learning about non-standard BBN scenarios and extra relativistic degrees of freedom. 

\begin{figure*}
    \centering
    \includegraphics[width=1.86\columnwidth]{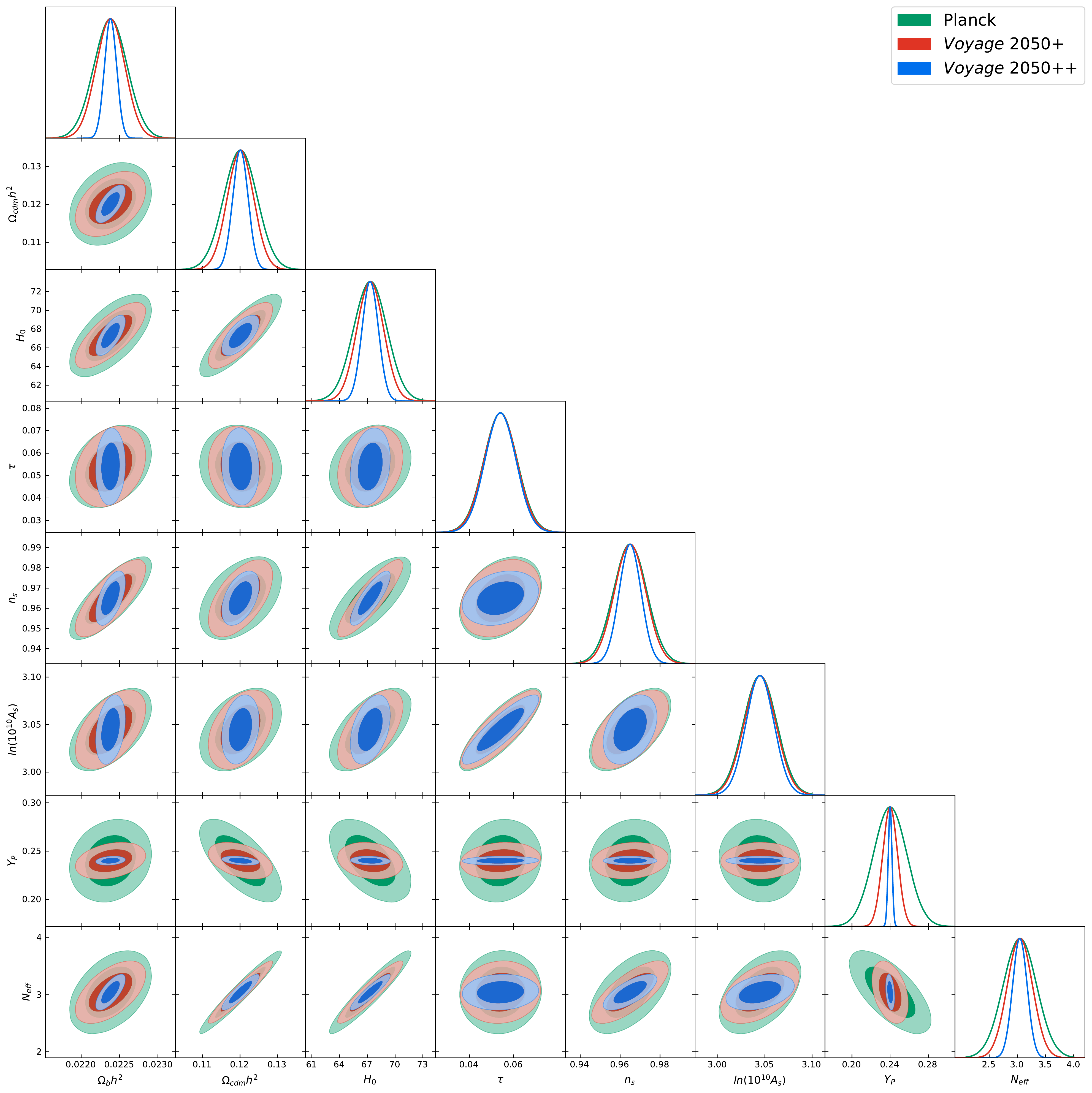}
    \caption{This figure depicts the improvement in parameter constraints over \Planck, estimated using a Fisher forecast for \Voyagep and a 5 times improved version of \Voyagep (labeled \Voyagepp) including astrophysical and distortion foregrounds.}
    \label{fig:contour_v2050X10}
\end{figure*}

In Fig.~\ref{fig:contour_v2050X10}, we also illustrate the covariances of all the parameters in the case with $\Yp$ and $\Neff$ added. The new constraints from the CRR enter through $\omb$ and $\Yp$ but then propagate to all the other parameters through the covariances in the \Planck data. This means that \Voyagep in combination with \Planck can also achieve marginal improvements on $\omc$ and $\Ho$ even if the CRR does not directly add significant information on these. Changes to $\ns$ and $\As$ are not visible for \Voyagep, but by further increasing the sensitivity by another factor of 5 we can even start seeing gains on $\ns$ (Fig.~\ref{fig:contour_v2050X10}).
These can reach a factor of $\simeq 1.5$ when comparing to \Planck alone. Similarly, geometric degeneracies can be further broken and improvements on $\Ho$ and $\omc$ by a factor of $\simeq 2$ can be achieved. 
This highlights the great potential of adding CRR constraints as a probe of fundamental physics, however, we are already in the regime of $\simeq 1500$ times the sensitivity of \PIXIE.

\vspace{-3mm}
\section{Conclusions}
\label{sec:conclusions}
The CRR provides a rich spectral signal that can be used to constrain cosmological parameters. 
In this work, we provided detailed forecasts illustrating the potential of future CRR measurements. 
We used state-of-the-art computations of the CRR and included the dominant foregrounds in our calculations. 
This allowed us to show that experimental concepts like \Voyage and \Voyagep can obtain significant detections of the CRR, while interesting constraints on cosmological parameters requires \Voyagep, which is essentially $\simeq 300$ times more sensitive than \PIXIE.

Without foregrounds, $\omb$ and $\Yp$ can in principle be measured to very high significance with both \Voyage and \Voyagep (Table~\ref{tab:Spec_case_II}). \Voyagep can furthermore in principle constrain $\omc$ if foregrounds (both astrophysical and from spectral distortions) were perfectly removed, while limits on $\Neff$ during recombination and $\Ho$ still remain weak. This can be understood by comparing the level of the overall CRR responses with respect to each of these parameters (Fig.~\ref{fig:parameters}).

The effect of $\omb$ is mainly a change in the amplitude of the CRR and the constraint is strongly weakened once foregrounds are included. In comparison to existing constraints from CMB anisotropies, from the CRR {\it alone} an improvement can only be achieved if the sensitivity of \Voyagep is further increased by a factor $\gtrsim 2-3$. To be on par with planned CMB experiments like SO \citep{SOWP2018}, a factor $\gtrsim 6$ over \Voyagep would be needed, making this perspective extremely futuristic. Independent gains on other parameters like $\omc$ and $\Neff$ during recombination remain challenging, even then.

However, the real potential lies in the sensitivity of the CRR to $\Yp$. Even in the presence of foregrounds \Voyagep could yield a $\simeq 4\%$ measurement of $\Yp$. The degradation by foregrounds is only a factor of $\simeq 2$, given the significant frequency dependence of the CRR response (Fig.~\ref{fig:parameters}). This alone is comparable to the limit from \Planck \citep{Planck2018params} and is a factor of $\simeq 1.7$ weaker than the constraint that SO could achieve. 
By further combining CRR measurements with \Planck, we find that the limits on $\Yp$ can be improved by a factor of $\simeq 2$, thus becoming comparable to those from SO. Marginal gains on $\omb$, $\omc$ and $\Neff$ can also be expected in this case (Fig.~\ref{fig:contour_v2050X10}).
The CRR has thus the potential to become a precision probe of non-standard BBN models \citep[e.g.,][]{Pospelov2010}.

While the CRR limit on $\Neff$ itself is very weak, one should emphasize the two important aspects. As shown here, the CRR response is actually sensitive to each of the three recombination eras (Fig.~\ref{fig:Neff_direct}), and thus in principle provides independent measurements of $\Neff$ at various redshifts during recombination.
On the other hand, by using BBN consistency relations one can indirectly obtain an interesting limit on $\NeffBBN$ around BBN when combined with \Planck. This is because the response is amplified through the modifications of $\Ypb(\Neff)$ (Sect.~\ref{sec:BBN}). 
The CRR thus provides another independent avenue for constraining the presence of extra relativistic degrees of freedom \citep[e.g., dark radiation,][]{Archidiacono2011, Menestrina2012} at various redshifts. However, to reach highly competitive limits one would probably need another factor of $\gtrsim 4$ in sensitivity over \Voyagep. 
In this case, one can also expect significant gains on all other standard cosmological parameters, including those relating to the primordial power spectrum (Fig.~\ref{fig:contour_v2050X10}). Thus the CRR could play an indirect role for understanding inflation physics and early-universe physics.

Given that the CRR provides a direct probe of the primordial helium abundance, it could be very interesting to consider inhomogeneous BBN scenarios \citep[e.g.,][]{Jedamzik2001, Nakamura2010} that may create anisotropies in $\Yp$ or more generally lead to deviant patches \citep{Chary2016, Arbey2020}. Through measurements of the CRR in different directions, these models could be constrained. In addition, if fluctuations in $\Yp$ are present at extremely small scales, this could potentially still be observable since at second order in $\Delta\Yp/\Yp$ or $\Delta \omb/\omb$ a non-negligible modification of the average CRR may be visible. While challenging, this could be another unique opportunity for future CMB spectrometers targeting the CRR.

In this work, we only considered fixed experimental concepts without trying to further optimize the distribution of channels or frequency resolution. This would certainly deserve additional investigation as it might be possible to reduce the frequency range over which the measurements are performed. Since the foreground penalties are not as severe for the CRR (only a factor of $\simeq 2$ for $\Yp$), we expect low-frequency channels to be less important than for measurements of $\mu$-distortions \citep{Abitbol2017}. This could help reduce the complexity and cost of the spectrometer. 
In addition, the observing strategy should be optimized. Here, we considered all-sky observation, but observations on single clean patches might be beneficial. The modeling of spatially-varying foregrounds should furthermore be considered more carefully, e.g., by using combinations of moment-expansion and ILC methods \citep{Rotti2020}.
We leave these tasks to future work.

We close by noting that at the level of \Voyage and \Voyagep, one also expects a significant detection of the primordial $\mu$ distortion from the dissipation of acoustic modes. In combination with SO this could yield improvements of the slow-roll parameters or simply provides independent measurements of the amplitude of the curvature power spectrum at wavenumbers $k\simeq 10^3-10^4\,\Mpc^{-1}$ \citep{Chluba2019Voyage}. The sensitivity of the $\mu$ constraint on the marginalization over foregrounds is significantly higher than for the CRR, simply because the $\mu$ distortion is so featureless. \SuperPIXIE and \Voyage would provide very important milestones for primordial distortion measurements, while \Voyagep would allow unprecedented constraints on inflation, both direct (through $\mu$) and indirect (through $\Yp$ and $\omb$ from the CRR).
We can only hope that these types of measurements will become possible in the distant future.

{\small {\it Acknowledgments}:
The authors would like to thank Josef Pradler for valuable discussion about inhomogeneous BBN scenarios. Fig.~\ref{fig:contour_v2050X10_spec} and \ref{fig:contour_v2050X10} were generated using the {\tt getdist} software package \citep{GetDist}.
This work was supported by the ERC Consolidator Grant {\it CMBSPEC} (No.~725456) as part of the European Union's Horizon 2020 research and innovation program.
JC was also supported by the Royal Society as a Royal Society URF at the University of Manchester.
}

\appendix

\bibliographystyle{mnras}
\bibliography{Lit}

\begin{thebibliography}{}
\makeatletter
\relax
\def\mn@urlcharsother{\let\do\@makeother \do\$\do\&\do\#\do\^\do\_\do\%\do\~}
\def\mn@doi{\begingroup\mn@urlcharsother \@ifnextchar [ {\mn@doi@}
  {\mn@doi@[]}}
\def\mn@doi@[#1]#2{\def\@tempa{#1}\ifx\@tempa\@empty \href
  {http://dx.doi.org/#2} {doi:#2}\else \href {http://dx.doi.org/#2} {#1}\fi
  \endgroup}
\def\mn@eprint#1#2{\mn@eprint@#1:#2::\@nil}
\def\mn@eprint@arXiv#1{\href {http://arxiv.org/abs/#1} {{\tt arXiv:#1}}}
\def\mn@eprint@dblp#1{\href {http://dblp.uni-trier.de/rec/bibtex/#1.xml}
  {dblp:#1}}
\def\mn@eprint@#1:#2:#3:#4\@nil{\def\@tempa {#1}\def\@tempb {#2}\def\@tempc
  {#3}\ifx \@tempc \@empty \let \@tempc \@tempb \let \@tempb \@tempa \fi \ifx
  \@tempb \@empty \def\@tempb {arXiv}\fi \@ifundefined
  {mn@eprint@\@tempb}{\@tempb:\@tempc}{\expandafter \expandafter \csname
  mn@eprint@\@tempb\endcsname \expandafter{\@tempc}}}

\bibitem[\protect\citeauthoryear{{Abazajian} et~al.,}{{Abazajian}
  et~al.}{2015}]{Abazajian2015}
{Abazajian} K.~N.,  et~al., 2015, \mn@doi [Astroparticle Physics]
  {10.1016/j.astropartphys.2014.05.014}, \href
  {http://adsabs.harvard.edu/abs/2015APh....63...66A} {63, 66}

\bibitem[\protect\citeauthoryear{Abitbol, Chluba, Hill  \& Johnson}{Abitbol
  et~al.}{2017}]{Abitbol2017}
Abitbol M.~H.,  Chluba J.,  Hill J.~C.,   Johnson B.~R.,  2017, \mn@doi
  [Monthly Notices of the Royal Astronomical Society] {10.1093/mnras/stx1653},
  471, 1126–1140

\bibitem[\protect\citeauthoryear{{Abitbol}, {Hill}  \& {Chluba}}{{Abitbol}
  et~al.}{2020}]{Abitbol2020}
{Abitbol} M.~H.,  {Hill} J.~C.,   {Chluba} J.,  2020, \mn@doi [\apj]
  {10.3847/1538-4357/ab7b70}, \href
  {https://ui.adsabs.harvard.edu/abs/2020ApJ...893...18A} {893, 18}

\bibitem[\protect\citeauthoryear{{Ade} et~al.,}{{Ade} et~al.}{2019}]{SOWP2018}
{Ade} P.,  et~al., 2019, \mn@doi [\jcap] {10.1088/1475-7516/2019/02/056}, \href
  {https://ui.adsabs.harvard.edu/abs/2019JCAP...02..056A} {2019, 056}

\bibitem[\protect\citeauthoryear{{Ali-Ha{\"i}moud}}{{Ali-Ha{\"i}moud}}{2013}]{Yacine2013RecSpec}
{Ali-Ha{\"i}moud} Y.,  2013, \mn@doi [\prd] {10.1103/PhysRevD.87.023526}, \href
  {http://adsabs.harvard.edu/abs/2013PhRvD..87b3526A} {87, 023526}

\bibitem[\protect\citeauthoryear{{Ali-Ha{\"i}moud} \&
  {Hirata}}{{Ali-Ha{\"i}moud} \& {Hirata}}{2010}]{Yacine2010}
{Ali-Ha{\"i}moud} Y.,  {Hirata} C.~M.,  2010, \mn@doi [\prd]
  {10.1103/PhysRevD.82.063521}, \href
  {http://adsabs.harvard.edu/abs/2010PhRvD..82f3521A} {82, 063521}

\bibitem[\protect\citeauthoryear{{Ali-Ha{\"i}moud} \&
  {Hirata}}{{Ali-Ha{\"i}moud} \& {Hirata}}{2011}]{Yacine2010c}
{Ali-Ha{\"i}moud} Y.,  {Hirata} C.~M.,  2011, \mn@doi [\prd]
  {10.1103/PhysRevD.83.043513}, \href
  {http://adsabs.harvard.edu/abs/2011PhRvD..83d3513A} {83, 043513}

\bibitem[\protect\citeauthoryear{{Ali-Ha{\"i}moud}, {Grin}  \&
  {Hirata}}{{Ali-Ha{\"i}moud} et~al.}{2010}]{Yacine2010b}
{Ali-Ha{\"i}moud} Y.,  {Grin} D.,   {Hirata} C.~M.,  2010, \mn@doi [\prd]
  {10.1103/PhysRevD.82.123502}, \href
  {http://adsabs.harvard.edu/abs/2010PhRvD..82l3502A} {82, 123502}

\bibitem[\protect\citeauthoryear{{Arbey}, {Auffinger}  \& {Silk}}{{Arbey}
  et~al.}{2020}]{Arbey2020}
{Arbey} A.,  {Auffinger} J.,   {Silk} J.,  2020, arXiv e-prints, \href
  {https://ui.adsabs.harvard.edu/abs/2020arXiv200602446A} {p. arXiv:2006.02446}

\bibitem[\protect\citeauthoryear{{Archidiacono}, {Calabrese}  \&
  {Melchiorri}}{{Archidiacono} et~al.}{2011}]{Archidiacono2011}
{Archidiacono} M.,  {Calabrese} E.,   {Melchiorri} A.~r.,  2011, \mn@doi [\prd]
  {10.1103/PhysRevD.84.123008}, \href
  {https://ui.adsabs.harvard.edu/abs/2011PhRvD..84l3008A} {84, 123008}

\bibitem[\protect\citeauthoryear{{Bennett} et~al.,}{{Bennett}
  et~al.}{1996}]{COBE4yr}
{Bennett} C.~L.,  et~al., 1996, \mn@doi [\apjl] {10.1086/310075}, \href
  {http://adsabs.harvard.edu/abs/1996ApJ...464L...1B} {464, L1}

\bibitem[\protect\citeauthoryear{{Burgin}}{{Burgin}}{2003}]{Burgin2003}
{Burgin} M.~S.,  2003, \mn@doi [Astronomy Reports] {10.1134/1.1611211}, \href
  {http://adsabs.harvard.edu/abs/2003ARep...47..709B} {47, 709}

\bibitem[\protect\citeauthoryear{{Burigana}, {Danese}  \& {de
  Zotti}}{{Burigana} et~al.}{1991}]{Burigana1991}
{Burigana} C.,  {Danese} L.,   {de Zotti} G.,  1991, \aap, \href
  {http://adsabs.harvard.edu/abs/1991A%26A...246...49B} {246, 49}

\bibitem[\protect\citeauthoryear{{Cabass}, {Melchiorri}  \& {Pajer}}{{Cabass}
  et~al.}{2016}]{Cabass2016}
{Cabass} G.,  {Melchiorri} A.,   {Pajer} E.,  2016, \mn@doi [\prd]
  {10.1103/PhysRevD.93.083515}, \href
  {http://adsabs.harvard.edu/abs/2016PhRvD..93h3515C} {93, 083515}

\bibitem[\protect\citeauthoryear{{Chary}}{{Chary}}{2016}]{Chary2016}
{Chary} R.,  2016, \mn@doi [\apj] {10.3847/0004-637X/817/1/33}, \href
  {https://ui.adsabs.harvard.edu/abs/2016ApJ...817...33C} {817, 33}

\bibitem[\protect\citeauthoryear{{Chluba}}{{Chluba}}{2005}]{Chluba2005}
{Chluba} J.,  2005, PhD thesis, LMU M{\"u}nchen

\bibitem[\protect\citeauthoryear{{Chluba}}{{Chluba}}{2010}]{Chluba2010a}
{Chluba} J.,  2010, \mn@doi [\mnras] {10.1111/j.1365-2966.2009.15957.x}, \href
  {http://adsabs.harvard.edu/abs/2010MNRAS.402.1195C} {402, 1195}

\bibitem[\protect\citeauthoryear{{Chluba}}{{Chluba}}{2013}]{Chluba2013fore}
{Chluba} J.,  2013, \mn@doi [\mnras] {10.1093/mnras/stt1733}, \href
  {http://adsabs.harvard.edu/abs/2013MNRAS.436.2232C} {436, 2232}

\bibitem[\protect\citeauthoryear{{Chluba}}{{Chluba}}{2014}]{Chluba2014TRR}
{Chluba} J.,  2014, \mn@doi [\mnras] {10.1093/mnras/stu1260}, \href
  {https://ui.adsabs.harvard.edu/abs/2014MNRAS.443.1881C} {443, 1881}

\bibitem[\protect\citeauthoryear{{Chluba}}{{Chluba}}{2015}]{Chluba2015GreensII}
{Chluba} J.,  2015, \mn@doi [\mnras] {10.1093/mnras/stv2243}, \href
  {http://adsabs.harvard.edu/abs/2015MNRAS.454.4182C} {454, 4182}

\bibitem[\protect\citeauthoryear{{Chluba}}{{Chluba}}{2016}]{Chluba2016}
{Chluba} J.,  2016, \mn@doi [\mnras] {10.1093/mnras/stw945}, \href
  {http://adsabs.harvard.edu/abs/2016MNRAS.460..227C} {460, 227}

\bibitem[\protect\citeauthoryear{{Chluba} \& {Ali-Ha{\"i}moud}}{{Chluba} \&
  {Ali-Ha{\"i}moud}}{2016}]{Chluba2016CosmoSpec}
{Chluba} J.,  {Ali-Ha{\"i}moud} Y.,  2016, \mn@doi [\mnras]
  {10.1093/mnras/stv2691}, \href
  {http://adsabs.harvard.edu/abs/2016MNRAS.456.3494C} {456, 3494}

\bibitem[\protect\citeauthoryear{{Chluba} \& {Jeong}}{{Chluba} \&
  {Jeong}}{2014}]{Chluba2013PCA}
{Chluba} J.,  {Jeong} D.,  2014, \mn@doi [\mnras] {10.1093/mnras/stt2327},
  \href {http://adsabs.harvard.edu/abs/2014MNRAS.438.2065C} {438, 2065}

\bibitem[\protect\citeauthoryear{{Chluba} \& {Sunyaev}}{{Chluba} \&
  {Sunyaev}}{2006a}]{Chluba2006}
{Chluba} J.,  {Sunyaev} R.~A.,  2006a, \mn@doi [\aap]
  {10.1051/0004-6361:20053988}, \href
  {http://cdsads.u-strasbg.fr/cgi-bin/nph-bib_query?bibcode=2006A%26A...446...39C&db_key=AST}
  {446, 39}

\bibitem[\protect\citeauthoryear{{Chluba} \& {Sunyaev}}{{Chluba} \&
  {Sunyaev}}{2006b}]{Chluba2006b}
{Chluba} J.,  {Sunyaev} R.~A.,  2006b, \mn@doi [\aap]
  {10.1051/0004-6361:20066191}, \href
  {http://cdsads.u-strasbg.fr/cgi-bin/nph-bib_query?bibcode=2006A%26A...458L..29C&db_key=AST}
  {458, L29}

\bibitem[\protect\citeauthoryear{{Chluba} \& {Sunyaev}}{{Chluba} \&
  {Sunyaev}}{2008a}]{Chluba2008T0}
{Chluba} J.,  {Sunyaev} R.~A.,  2008a, \mn@doi [\aap]
  {10.1051/0004-6361:20078200}, \href
  {http://adsabs.harvard.edu/abs/2008A%26A...478L..27C} {478, L27}

\bibitem[\protect\citeauthoryear{{Chluba} \& {Sunyaev}}{{Chluba} \&
  {Sunyaev}}{2008b}]{Chluba2008a}
{Chluba} J.,  {Sunyaev} R.~A.,  2008b, \mn@doi [\aap]
  {10.1051/0004-6361:20077921}, \href
  {http://adsabs.harvard.edu/abs/2008A%26A...480..629C} {480, 629}

\bibitem[\protect\citeauthoryear{{Chluba} \& {Sunyaev}}{{Chluba} \&
  {Sunyaev}}{2009a}]{Chluba2008b}
{Chluba} J.,  {Sunyaev} R.~A.,  2009a, \mn@doi [\aap]
  {10.1051/0004-6361/200811100}, \href
  {http://adsabs.harvard.edu/abs/2009A%26A...496..619C} {496, 619}

\bibitem[\protect\citeauthoryear{{Chluba} \& {Sunyaev}}{{Chluba} \&
  {Sunyaev}}{2009b}]{Chluba2008c}
{Chluba} J.,  {Sunyaev} R.~A.,  2009b, \mn@doi [\aap]
  {10.1051/0004-6361/200809840}, \href
  {http://adsabs.harvard.edu/abs/2009A%26A...501...29C} {501, 29}

\bibitem[\protect\citeauthoryear{{Chluba} \& {Sunyaev}}{{Chluba} \&
  {Sunyaev}}{2010a}]{Chluba2009c}
{Chluba} J.,  {Sunyaev} R.~A.,  2010a, \mn@doi [\mnras]
  {10.1111/j.1365-2966.2009.15959.x}, \href
  {http://adsabs.harvard.edu/abs/2010MNRAS.402.1221C} {402, 1221}

\bibitem[\protect\citeauthoryear{{Chluba} \& {Sunyaev}}{{Chluba} \&
  {Sunyaev}}{2010b}]{Chluba2009}
{Chluba} J.,  {Sunyaev} R.~A.,  2010b, \mn@doi [\aap]
  {10.1051/0004-6361/200912263}, \href
  {http://adsabs.harvard.edu/abs/2010A%26A...512A..53C} {512, A53+}

\bibitem[\protect\citeauthoryear{{Chluba} \& {Sunyaev}}{{Chluba} \&
  {Sunyaev}}{2012}]{Chluba2011therm}
{Chluba} J.,  {Sunyaev} R.~A.,  2012, \mn@doi [\mnras]
  {10.1111/j.1365-2966.2011.19786.x}, \href
  {http://adsabs.harvard.edu/abs/2012MNRAS.419.1294C} {419, 1294}

\bibitem[\protect\citeauthoryear{{Chluba} \& {Thomas}}{{Chluba} \&
  {Thomas}}{2011}]{Chluba2010b}
{Chluba} J.,  {Thomas} R.~M.,  2011, \mn@doi [\mnras]
  {10.1111/j.1365-2966.2010.17940.x}, \href
  {http://adsabs.harvard.edu/abs/2011MNRAS.412..748C} {412, 748}

\bibitem[\protect\citeauthoryear{{Chluba}, {Rubi{\~n}o-Mart{\'{\i}}n}  \&
  {Sunyaev}}{{Chluba} et~al.}{2007}]{Chluba2007}
{Chluba} J.,  {Rubi{\~n}o-Mart{\'{\i}}n} J.~A.,   {Sunyaev} R.~A.,  2007,
  \mn@doi [\mnras] {10.1111/j.1365-2966.2006.11239.x}, \href
  {http://cdsads.u-strasbg.fr/cgi-bin/nph-bib_query?bibcode=2007MNRAS.374.1310C&db_key=AST}
  {374, 1310}

\bibitem[\protect\citeauthoryear{{Chluba}, {Vasil}  \& {Dursi}}{{Chluba}
  et~al.}{2010}]{Chluba2010}
{Chluba} J.,  {Vasil} G.~M.,   {Dursi} L.~J.,  2010, \mn@doi [\mnras]
  {10.1111/j.1365-2966.2010.16940.x}, \href
  {http://adsabs.harvard.edu/abs/2010MNRAS.407..599C} {407, 599}

\bibitem[\protect\citeauthoryear{{Chluba}, {Fung}  \& {Switzer}}{{Chluba}
  et~al.}{2012a}]{Chluba2012HeRec}
{Chluba} J.,  {Fung} J.,   {Switzer} E.~R.,  2012a, \mn@doi [\mnras]
  {10.1111/j.1365-2966.2012.21110.x}, \href
  {http://adsabs.harvard.edu/abs/2012MNRAS.423.3227C} {423, 3227}

\bibitem[\protect\citeauthoryear{{Chluba}, {Khatri}  \& {Sunyaev}}{{Chluba}
  et~al.}{2012b}]{Chluba2012}
{Chluba} J.,  {Khatri} R.,   {Sunyaev} R.~A.,  2012b, \mn@doi [\mnras]
  {10.1111/j.1365-2966.2012.21474.x}, \href
  {http://adsabs.harvard.edu/abs/2012MNRAS.425.1129C} {425, 1129}

\bibitem[\protect\citeauthoryear{{Chluba}, {Erickcek}  \& {Ben-Dayan}}{{Chluba}
  et~al.}{2012c}]{Chluba2012inflaton}
{Chluba} J.,  {Erickcek} A.~L.,   {Ben-Dayan} I.,  2012c, \mn@doi [\apj]
  {10.1088/0004-637X/758/2/76}, \href
  {http://adsabs.harvard.edu/abs/2012ApJ...758...76C} {758, 76}

\bibitem[\protect\citeauthoryear{{Chluba}, {Hill}  \& {Abitbol}}{{Chluba}
  et~al.}{2017}]{Chluba2017}
{Chluba} J.,  {Hill} J.~C.,   {Abitbol} M.~H.,  2017, ArXiv:1701.00274, \href
  {http://adsabs.harvard.edu/abs/2017arXiv170100274C} {}

\bibitem[\protect\citeauthoryear{{Chluba} et~al.,}{{Chluba}
  et~al.}{2019a}]{Chluba2019Voyage}
{Chluba} J.,  et~al., 2019a, arXiv e-prints, \href
  {https://ui.adsabs.harvard.edu/abs/2019arXiv190901593C} {p. arXiv:1909.01593}

\bibitem[\protect\citeauthoryear{{Chluba} et~al.,}{{Chluba}
  et~al.}{2019b}]{Chluba2019BAAS}
{Chluba} J.,  et~al., 2019b, BAAS, \href
  {https://ui.adsabs.harvard.edu/abs/2019BAAS...51c.184C} {51, 184}

\bibitem[\protect\citeauthoryear{{Daly}}{{Daly}}{1991}]{Daly1991}
{Daly} R.~A.,  1991, \mn@doi [\apj] {10.1086/169866}, \href
  {http://adsabs.harvard.edu/abs/1991ApJ...371...14D} {371, 14}

\bibitem[\protect\citeauthoryear{{Danese} \& {de Zotti}}{{Danese} \& {de
  Zotti}}{1981}]{Danese1981}
{Danese} L.,  {de Zotti} G.,  1981, \aap, \href
  {http://adsabs.harvard.edu/abs/1981A%26A....94L..33D} {94, L33}

\bibitem[\protect\citeauthoryear{{Desjacques}, {Chluba}, {Silk}, {de Bernardis}
   \& {Dor{\'e}}}{{Desjacques} et~al.}{2015}]{Vince2015}
{Desjacques} V.,  {Chluba} J.,  {Silk} J.,  {de Bernardis} F.,   {Dor{\'e}} O.,
   2015, \mn@doi [\mnras] {10.1093/mnras/stv1291}, \href
  {http://adsabs.harvard.edu/abs/2015MNRAS.451.4460D} {451, 4460}

\bibitem[\protect\citeauthoryear{{Dubrovich}}{{Dubrovich}}{1975}]{Dubrovich1975}
{Dubrovich} V.~K.,  1975, Soviet Astronomy Letters, \href
  {http://cdsads.u-strasbg.fr/cgi-bin/nph-bib_query?bibcode=1975SvAL....1..196D&db_key=AST}
  {1, 196}

\bibitem[\protect\citeauthoryear{{Dubrovich} \& {Stolyarov}}{{Dubrovich} \&
  {Stolyarov}}{1995}]{DubroVlad95}
{Dubrovich} V.~K.,  {Stolyarov} V.~A.,  1995, \aap, \href
  {http://adsabs.harvard.edu/cgi-bin/nph-bib_query?bibcode=1995A%26A...302..635D&db_key=AST}
  {302, 635}

\bibitem[\protect\citeauthoryear{{Dubrovich} \& {Stolyarov}}{{Dubrovich} \&
  {Stolyarov}}{1997}]{Dubrovich1997}
{Dubrovich} V.~K.,  {Stolyarov} V.~A.,  1997, Astronomy Letters, \href
  {http://adsabs.harvard.edu/abs/1997AstL...23..565D} {23, 565}

\bibitem[\protect\citeauthoryear{{Fendt}}{{Fendt}}{2009}]{Fendt2009Thesis}
{Fendt} William~Ashton J.,  2009, PhD thesis, University of Illinois at
  Urbana-Champaign

\bibitem[\protect\citeauthoryear{{Fendt}, {Chluba}, {Rubi{\~n}o-Mart{\'{\i}}n}
  \& {Wandelt}}{{Fendt} et~al.}{2009}]{Fendt2009}
{Fendt} W.~A.,  {Chluba} J.,  {Rubi{\~n}o-Mart{\'{\i}}n} J.~A.,   {Wandelt}
  B.~D.,  2009, \mn@doi [\apjs] {10.1088/0067-0049/181/2/627}, \href
  {http://adsabs.harvard.edu/abs/2009ApJS..181..627F} {181, 627}

\bibitem[\protect\citeauthoryear{{Fixsen}}{{Fixsen}}{2009}]{Fixsen2009}
{Fixsen} D.~J.,  2009, \mn@doi [\apj] {10.1088/0004-637X/707/2/916}, \href
  {http://adsabs.harvard.edu/abs/2009ApJ...707..916F} {707, 916}

\bibitem[\protect\citeauthoryear{{Fixsen} \& {Mather}}{{Fixsen} \&
  {Mather}}{2002}]{Fixsen2002}
{Fixsen} D.~J.,  {Mather} J.~C.,  2002, \mn@doi [\apj] {10.1086/344402}, \href
  {http://adsabs.harvard.edu/cgi-bin/nph-bib_query?bibcode=2002ApJ...581..817F&db_key=AST}
  {581, 817}

\bibitem[\protect\citeauthoryear{{Fixsen}, {Cheng}, {Gales}, {Mather}, {Shafer}
   \& {Wright}}{{Fixsen} et~al.}{1996}]{Fixsen1996}
{Fixsen} D.~J.,  {Cheng} E.~S.,  {Gales} J.~M.,  {Mather} J.~C.,  {Shafer}
  R.~A.,   {Wright} E.~L.,  1996, \mn@doi [\apj] {10.1086/178173}, \href
  {http://adsabs.harvard.edu/abs/1996ApJ...473..576F} {473, 576}

\bibitem[\protect\citeauthoryear{{Fixsen} et~al.,}{{Fixsen}
  et~al.}{2011}]{Fixsen2011}
{Fixsen} D.~J.,  et~al., 2011, \mn@doi [\apj] {10.1088/0004-637X/734/1/5},
  \href {http://adsabs.harvard.edu/abs/2011ApJ...734....5F} {734, 5}

\bibitem[\protect\citeauthoryear{{Grin} \& {Hirata}}{{Grin} \&
  {Hirata}}{2010}]{Grin2009}
{Grin} D.,  {Hirata} C.~M.,  2010, \mn@doi [\prd] {10.1103/PhysRevD.81.083005},
  \href {http://adsabs.harvard.edu/abs/2010PhRvD..81h3005G} {81, 083005}

\bibitem[\protect\citeauthoryear{Hart \& Chluba}{Hart \&
  Chluba}{2018}]{Hart2017}
Hart L.,  Chluba J.,  2018, \mn@doi [MNRAS] {10.1093/mnras/stx2783}, 474, 1850

\bibitem[\protect\citeauthoryear{{Hill}, {Battaglia}, {Chluba}, {Ferraro},
  {Schaan}  \& {Spergel}}{{Hill} et~al.}{2015}]{Hill2015}
{Hill} J.~C.,  {Battaglia} N.,  {Chluba} J.,  {Ferraro} S.,  {Schaan} E.,
  {Spergel} D.~N.,  2015, \mn@doi [Physical Review Letters]
  {10.1103/PhysRevLett.115.261301}, \href
  {http://adsabs.harvard.edu/abs/2015PhRvL.115z1301H} {115, 261301}

\bibitem[\protect\citeauthoryear{{Hinshaw} et~al.,}{{Hinshaw}
  et~al.}{2013}]{wmap9params}
{Hinshaw} G.,  et~al., 2013, \mn@doi [\apjs] {10.1088/0067-0049/208/2/19},
  \href {http://adsabs.harvard.edu/abs/2013ApJS..208...19H} {208, 19}

\bibitem[\protect\citeauthoryear{{Hirata}}{{Hirata}}{2008}]{Hirata2008}
{Hirata} C.~M.,  2008, \mn@doi [\prd] {10.1103/PhysRevD.78.023001}, \href
  {http://adsabs.harvard.edu/abs/2008PhRvD..78b3001H} {78, 023001}

\bibitem[\protect\citeauthoryear{{Hirata} \& {Switzer}}{{Hirata} \&
  {Switzer}}{2008}]{Switzer2007III}
{Hirata} C.~M.,  {Switzer} E.~R.,  2008, \mn@doi [\prd]
  {10.1103/PhysRevD.77.083007}, \href
  {http://adsabs.harvard.edu/abs/2008PhRvD..77h3007H} {77, 083007}

\bibitem[\protect\citeauthoryear{Hou, Keisler, Knox, Millea  \& Reichardt}{Hou
  et~al.}{2013}]{Hou2013}
Hou Z.,  Keisler R.,  Knox L.,  Millea M.,   Reichardt C.,  2013, \mn@doi
  [Physical Review D] {10.1103/physrevd.87.083008}, 87

\bibitem[\protect\citeauthoryear{{Hu} \& {Silk}}{{Hu} \& {Silk}}{1993}]{Hu1993}
{Hu} W.,  {Silk} J.,  1993, \mn@doi [\prd] {10.1103/PhysRevD.48.485}, \href
  {http://adsabs.harvard.edu/abs/1993PhRvD..48..485H} {48, 485}

\bibitem[\protect\citeauthoryear{{Hu}, {Scott}  \& {Silk}}{{Hu}
  et~al.}{1994}]{Hu1994}
{Hu} W.,  {Scott} D.,   {Silk} J.,  1994, \mn@doi [\apjl] {10.1086/187424},
  \href {http://adsabs.harvard.edu/abs/1994ApJ...430L...5H} {430, L5}

\bibitem[\protect\citeauthoryear{{Illarionov} \& {Sunyaev}}{{Illarionov} \&
  {Sunyaev}}{1974}]{Illarionov1974}
{Illarionov} A.~F.,  {Sunyaev} R.~A.,  1974, Astronomicheskii Zhurnal, \href
  {http://adsabs.harvard.edu/abs/1974AZh....51.1162I} {51, 1162}

\bibitem[\protect\citeauthoryear{{Inogamov} \& {Sunyaev}}{{Inogamov} \&
  {Sunyaev}}{2015}]{Inogamov2015}
{Inogamov} N.~A.,  {Sunyaev} R.~A.,  2015, \mn@doi [Astronomy Letters]
  {10.1134/S106377371512004X}, \href
  {https://ui.adsabs.harvard.edu/abs/2015AstL...41..693I} {41, 693}

\bibitem[\protect\citeauthoryear{{Iocco}, {Mangano}, {Miele}, {Pisanti}  \&
  {Serpico}}{{Iocco} et~al.}{2009}]{Iocco2009}
{Iocco} F.,  {Mangano} G.,  {Miele} G.,  {Pisanti} O.,   {Serpico} P.~D.,
  2009, \mn@doi [Physics Reports] {10.1016/j.physrep.2009.02.002}, \href
  {http://adsabs.harvard.edu/abs/2009PhR...472....1I} {472, 1}

\bibitem[\protect\citeauthoryear{{Ivanov}, {Ali-Ha{\"\i}moud}  \&
  {Lesgourgues}}{{Ivanov} et~al.}{2020}]{Ivanov2020}
{Ivanov} M.~M.,  {Ali-Ha{\"\i}moud} Y.,   {Lesgourgues} J.,  2020, arXiv
  e-prints, \href {https://ui.adsabs.harvard.edu/abs/2020arXiv200510656I} {p.
  arXiv:2005.10656}

\bibitem[\protect\citeauthoryear{{Jedamzik} \& {Rehm}}{{Jedamzik} \&
  {Rehm}}{2001}]{Jedamzik2001}
{Jedamzik} K.,  {Rehm} J.~B.,  2001, \mn@doi [\prd]
  {10.1103/PhysRevD.64.023510}, \href
  {https://ui.adsabs.harvard.edu/abs/2001PhRvD..64b3510J} {64, 023510}

\bibitem[\protect\citeauthoryear{{Khatri} \& {Sunyaev}}{{Khatri} \&
  {Sunyaev}}{2013}]{Khatri2013forecast}
{Khatri} R.,  {Sunyaev} R.~A.,  2013, \mn@doi [\jcap]
  {10.1088/1475-7516/2013/06/026}, \href
  {http://adsabs.harvard.edu/abs/2013JCAP...06..026K} {6, 26}

\bibitem[\protect\citeauthoryear{{Kholupenko} \& {Ivanchik}}{{Kholupenko} \&
  {Ivanchik}}{2006}]{Kholu2006}
{Kholupenko} E.~E.,  {Ivanchik} A.~V.,  2006, \mn@doi [Astronomy Letters]
  {10.1134/S1063773706120012}, \href
  {http://cdsads.u-strasbg.fr/cgi-bin/nph-bib_query?bibcode=2006AstL...32..795K&db_key=AST}
  {32, 795}

\bibitem[\protect\citeauthoryear{Kholupenko, Ivanchik  \&
  Varshalovich}{Kholupenko et~al.}{2005}]{Kholu2005}
Kholupenko E.~E.,  Ivanchik A.~V.,   Varshalovich D.~A.,  2005, Gravitation and
  Cosmology, 11, 161

\bibitem[\protect\citeauthoryear{{Kholupenko}, {Ivanchik}  \&
  {Varshalovich}}{{Kholupenko} et~al.}{2007}]{Kholupenko2007}
{Kholupenko} E.~E.,  {Ivanchik} A.~V.,   {Varshalovich} D.~A.,  2007, \mn@doi
  [\mnras] {10.1111/j.1745-3933.2007.00316.x}, \href
  {http://adsabs.harvard.edu/abs/2007MNRAS.378L..39K} {378, L39}

\bibitem[\protect\citeauthoryear{{Kogut} \& {Fixsen}}{{Kogut} \&
  {Fixsen}}{2018}]{Kogut2018beams}
{Kogut} A.~J.,  {Fixsen} D.~J.,  2018, \mn@doi [Journal of Astronomical
  Telescopes, Instruments, and Systems] {10.1117/1.JATIS.4.1.014006}, \href
  {https://ui.adsabs.harvard.edu/abs/2018JATIS...4a4006K} {4, 014006}

\bibitem[\protect\citeauthoryear{{Kogut} \& {Fixsen}}{{Kogut} \&
  {Fixsen}}{2019}]{Kogut2019System}
{Kogut} A.~J.,  {Fixsen} D.~J.,  2019, \mn@doi [Journal of Astronomical
  Telescopes, Instruments, and Systems] {10.1117/1.JATIS.5.2.024008}, \href
  {https://ui.adsabs.harvard.edu/abs/2019JATIS...5b4008K} {5, 024008}

\bibitem[\protect\citeauthoryear{{Kogut} \& {Fixsen}}{{Kogut} \&
  {Fixsen}}{2020}]{Kogut2020Cali}
{Kogut} A.,  {Fixsen} D.~J.,  2020, \mn@doi [\jcap]
  {10.1088/1475-7516/2020/05/041}, \href
  {https://ui.adsabs.harvard.edu/abs/2020JCAP...05..041K} {2020, 041}

\bibitem[\protect\citeauthoryear{{Kogut} et~al.,}{{Kogut}
  et~al.}{2011}]{Kogut2011PIXIE}
{Kogut} A.,  et~al., 2011, \mn@doi [\jcap] {10.1088/1475-7516/2011/07/025},
  \href {http://adsabs.harvard.edu/abs/2011JCAP...07..025K} {7, 25}

\bibitem[\protect\citeauthoryear{{Kogut}, {Chluba}, {Fixsen}, {Meyer}  \&
  {Spergel}}{{Kogut} et~al.}{2016}]{Kogut2016SPIE}
{Kogut} A.,  {Chluba} J.,  {Fixsen} D.~J.,  {Meyer} S.,   {Spergel} D.,  2016,
  in SPIE Conference Series. p. 99040W, \mn@doi{10.1117/12.2231090}

\bibitem[\protect\citeauthoryear{{Kogut}, {Abitbol}, {Chluba}, {Delabrouille},
  {Fixsen}, {Hill}, {Patil}  \& {Rotti}}{{Kogut} et~al.}{2019}]{Kogut2019BAAS}
{Kogut} A.,  {Abitbol} M.~H.,  {Chluba} J.,  {Delabrouille} J.,  {Fixsen} D.,
  {Hill} J.~C.,  {Patil} S.~P.,   {Rotti} A.,  2019, in BAAS. p.~113
  (\mn@eprint {arXiv} {1907.13195})

\bibitem[\protect\citeauthoryear{{Labzowsky}, {Shonin}  \&
  {Solovyev}}{{Labzowsky} et~al.}{2005}]{Labzowsky2005}
{Labzowsky} L.~N.,  {Shonin} A.~V.,   {Solovyev} D.~A.,  2005, \mn@doi [Journal
  of Physics B Atomic Molecular Physics] {10.1088/0953-4075/38/3/010}, \href
  {http://adsabs.harvard.edu/abs/2005JPhB...38..265L} {38, 265}

\bibitem[\protect\citeauthoryear{{Lee}, {Chluba}, {Kay}  \& {Barnes}}{{Lee}
  et~al.}{2020}]{Lee2020}
{Lee} E.,  {Chluba} J.,  {Kay} S.~T.,   {Barnes} D.~J.,  2020, \mn@doi [\mnras]
  {10.1093/mnras/staa450}, \href
  {https://ui.adsabs.harvard.edu/abs/2020MNRAS.493.3274L} {493, 3274}

\bibitem[\protect\citeauthoryear{Lewis}{Lewis}{2019}]{GetDist}
Lewis A.,  2019, GetDist: a Python package for analysing Monte Carlo samples
  (\mn@eprint {arXiv} {1910.13970})

\bibitem[\protect\citeauthoryear{Lewis \& Bridle}{Lewis \&
  Bridle}{2002}]{COSMOMC}
Lewis A.,  Bridle S.,  2002, Phys. Rev., D66, 103511

\bibitem[\protect\citeauthoryear{{Lucca}, {Sch{\"o}neberg}, {Hooper},
  {Lesgourgues}  \& {Chluba}}{{Lucca} et~al.}{2020}]{Lucca2020}
{Lucca} M.,  {Sch{\"o}neberg} N.,  {Hooper} D.~C.,  {Lesgourgues} J.,
  {Chluba} J.,  2020, \mn@doi [\jcap] {10.1088/1475-7516/2020/02/026}, \href
  {https://ui.adsabs.harvard.edu/abs/2020JCAP...02..026L} {2020, 026}

\bibitem[\protect\citeauthoryear{{Mather} et~al.,}{{Mather}
  et~al.}{1994}]{Mather1994}
{Mather} J.~C.,  et~al., 1994, \mn@doi [\apj] {10.1086/173574}, \href
  {http://adsabs.harvard.edu/abs/1994ApJ...420..439M} {420, 439}

\bibitem[\protect\citeauthoryear{{Menestrina} \& {Scherrer}}{{Menestrina} \&
  {Scherrer}}{2012}]{Menestrina2012}
{Menestrina} J.~L.,  {Scherrer} R.~J.,  2012, \mn@doi [\prd]
  {10.1103/PhysRevD.85.047301}, \href
  {https://ui.adsabs.harvard.edu/abs/2012PhRvD..85d7301M} {85, 047301}

\bibitem[\protect\citeauthoryear{{Mukhanov}, {Kim}, {Naselsky}, {Trombetti}  \&
  {Burigana}}{{Mukhanov} et~al.}{2012}]{Mukhanov2012}
{Mukhanov} V.,  {Kim} J.,  {Naselsky} P.,  {Trombetti} T.,   {Burigana} C.,
  2012, \mn@doi [\jcap] {10.1088/1475-7516/2012/06/040}, \href
  {https://ui.adsabs.harvard.edu/abs/2012JCAP...06..040M} {2012, 040}

\bibitem[\protect\citeauthoryear{{Mukherjee}, {Silk}  \& {Wandelt}}{{Mukherjee}
  et~al.}{2019}]{Mukherjee2019}
{Mukherjee} S.,  {Silk} J.,   {Wandelt} B.~D.,  2019, \mn@doi [\prd]
  {10.1103/PhysRevD.100.103508}, \href
  {https://ui.adsabs.harvard.edu/abs/2019PhRvD.100j3508M} {100, 103508}

\bibitem[\protect\citeauthoryear{{Nakamura}, {Hashimoto}, {Fujimoto},
  {Nishimura}  \& {Sato}}{{Nakamura} et~al.}{2010}]{Nakamura2010}
{Nakamura} R.,  {Hashimoto} M.-a.,  {Fujimoto} S.-i.,  {Nishimura} N.,   {Sato}
  K.,  2010, arXiv e-prints, \href
  {https://ui.adsabs.harvard.edu/abs/2010arXiv1007.0466N} {p. arXiv:1007.0466}

\bibitem[\protect\citeauthoryear{{PRISM Collaboration} et~al.,}{{PRISM
  Collaboration} et~al.}{2014}]{PRISM2013WPII}
{PRISM Collaboration} et~al., 2014, \mn@doi [\jcap]
  {10.1088/1475-7516/2014/02/006}, \href
  {http://adsabs.harvard.edu/abs/2014JCAP...02..006A} {2, 6}

\bibitem[\protect\citeauthoryear{{Pajer} \& {Zaldarriaga}}{{Pajer} \&
  {Zaldarriaga}}{2013}]{Pajer2012b}
{Pajer} E.,  {Zaldarriaga} M.,  2013, \mn@doi [\jcap]
  {10.1088/1475-7516/2013/02/036}, \href
  {http://adsabs.harvard.edu/abs/2013JCAP...02..036P} {2, 36}

\bibitem[\protect\citeauthoryear{{Peebles}}{{Peebles}}{1968}]{Peebles68}
{Peebles} P.~J.~E.,  1968, \apj, \href
  {http://adsabs.harvard.edu/cgi-bin/nph-bib_query?bibcode=1968ApJ...153....1P&db_key=AST}
  {153, 1}

\bibitem[\protect\citeauthoryear{Pisanti, Cirillo, Esposito, Iocco, Mangano,
  Miele  \& Serpico}{Pisanti et~al.}{2008}]{PARTHENOPE}
Pisanti O.,  Cirillo A.,  Esposito S.,  Iocco F.,  Mangano G.,  Miele G.,
  Serpico P.,  2008, \mn@doi [Computer Physics Communications]
  {10.1016/j.cpc.2008.02.015}, 178, 956–971

\bibitem[\protect\citeauthoryear{{Planck Collaboration et al.}}{{Planck
  Collaboration et al.}}{2015}]{Planck2015params}
{Planck Collaboration et al.} 2015, ArXiv:1502.01589

\bibitem[\protect\citeauthoryear{{Planck Collaboration et al.}}{{Planck
  Collaboration et al.}}{2018a}]{Planck2018params}
{Planck Collaboration et al.} 2018a, ArXiv:1807.06209

\bibitem[\protect\citeauthoryear{{Planck Collaboration et al.}}{{Planck
  Collaboration et al.}}{2018b}]{Planck2018over}
{Planck Collaboration et al.} 2018b, arXiv e-prints, \href
  {https://ui.adsabs.harvard.edu/abs/2018arXiv180706205P} {p. arXiv:1807.06205}

\bibitem[\protect\citeauthoryear{{Planck Collaboration et al.}}{{Planck
  Collaboration et al.}}{2019}]{Planck2018like}
{Planck Collaboration et al.} 2019, arXiv e-prints, \href
  {https://ui.adsabs.harvard.edu/abs/2019arXiv190712875P} {p. arXiv:1907.12875}

\bibitem[\protect\citeauthoryear{{Pospelov} \& {Pradler}}{{Pospelov} \&
  {Pradler}}{2010}]{Pospelov2010}
{Pospelov} M.,  {Pradler} J.,  2010, \mn@doi [Annual Review of Nuclear and
  Particle Science] {10.1146/annurev.nucl.012809.104521}, \href
  {http://adsabs.harvard.edu/abs/2010ARNPS..60..539P} {60, 539}

\bibitem[\protect\citeauthoryear{{Refregier} et~al.}{{Refregier}
  et~al.}{2000}]{Refregier2000}
{Refregier} A.,  et~al., 2000, \mn@doi [\prd] {10.1103/PhysRevD.61.123001},
  \href {http://adsabs.harvard.edu/abs/2000PhRvD..61l3001R} {61, 123001}

\bibitem[\protect\citeauthoryear{{Remazeilles}, {Bolliet}, {Rotti}  \&
  {Chluba}}{{Remazeilles} et~al.}{2019}]{Remazeilles2019}
{Remazeilles} M.,  {Bolliet} B.,  {Rotti} A.,   {Chluba} J.,  2019, \mn@doi
  [\mnras] {10.1093/mnras/sty3352}, \href
  {https://ui.adsabs.harvard.edu/abs/2019MNRAS.483.3459R} {483, 3459}

\bibitem[\protect\citeauthoryear{{Rotti} \& {Chluba}}{{Rotti} \&
  {Chluba}}{2020}]{Rotti2020}
{Rotti} A.,  {Chluba} J.,  2020, arXiv e-prints, \href
  {https://ui.adsabs.harvard.edu/abs/2020arXiv200602458R} {p. arXiv:2006.02458}

\bibitem[\protect\citeauthoryear{{Rubi{\~n}o-Mart{\'{\i}}n}, {Chluba}  \&
  {Sunyaev}}{{Rubi{\~n}o-Mart{\'{\i}}n} et~al.}{2006}]{Jose2006}
{Rubi{\~n}o-Mart{\'{\i}}n} J.~A.,  {Chluba} J.,   {Sunyaev} R.~A.,  2006,
  \mn@doi [\mnras] {10.1111/j.1365-2966.2006.10839.x}, \href
  {http://cdsads.u-strasbg.fr/cgi-bin/nph-bib_query?bibcode=2006MNRAS.371.1939R&db_key=AST}
  {371, 1939}

\bibitem[\protect\citeauthoryear{{Rubi{\~n}o-Mart{\'{\i}}n}, {Chluba}  \&
  {Sunyaev}}{{Rubi{\~n}o-Mart{\'{\i}}n} et~al.}{2008}]{Jose2008}
{Rubi{\~n}o-Mart{\'{\i}}n} J.~A.,  {Chluba} J.,   {Sunyaev} R.~A.,  2008,
  \mn@doi [\aap] {10.1051/0004-6361:20078993}, \href
  {http://adsabs.harvard.edu/abs/2008A%26A...485..377R} {485, 377}

\bibitem[\protect\citeauthoryear{{Rubi{\~n}o-Mart{\'{\i}}n}, {Chluba}, {Fendt}
  \& {Wandelt}}{{Rubi{\~n}o-Mart{\'{\i}}n} et~al.}{2010}]{Jose2010}
{Rubi{\~n}o-Mart{\'{\i}}n} J.~A.,  {Chluba} J.,  {Fendt} W.~A.,   {Wandelt}
  B.~D.,  2010, \mn@doi [\mnras] {10.1111/j.1365-2966.2009.16136.x}, \href
  {http://adsabs.harvard.edu/abs/2010MNRAS.403..439R} {403, 439}

\bibitem[\protect\citeauthoryear{{Rybicki} \& {dell'Antonio}}{{Rybicki} \&
  {dell'Antonio}}{1994}]{RybickiDell94}
{Rybicki} G.~B.,  {dell'Antonio} I.~P.,  1994, \mn@doi [\apj] {10.1086/174170},
  \href
  {http://cdsads.u-strasbg.fr/cgi-bin/nph-bib_query?bibcode=1994ApJ...427..603R&db_key=AST}
  {427, 603}

\bibitem[\protect\citeauthoryear{Sarkar \& Khatri}{Sarkar \&
  Khatri}{2020}]{Sarkar2020}
Sarkar D.,  Khatri R.,  2020, \mn@doi [\jcap] {10.1088/1475-7516/2020/01/050},
  2020, 050

\bibitem[\protect\citeauthoryear{{Sathyanarayana Rao}, {Subrahmanyan}, {Udaya
  Shankar}  \& {Chluba}}{{Sathyanarayana Rao} et~al.}{2015}]{Mayuri2015}
{Sathyanarayana Rao} M.,  {Subrahmanyan} R.,  {Udaya Shankar} N.,   {Chluba}
  J.,  2015, \mn@doi [\apj] {10.1088/0004-637X/810/1/3}, \href
  {http://adsabs.harvard.edu/abs/2015ApJ...810....3S} {810, 3}

\bibitem[\protect\citeauthoryear{{Shaw} \& {Chluba}}{{Shaw} \&
  {Chluba}}{2011}]{Shaw2011}
{Shaw} J.~R.,  {Chluba} J.,  2011, \mn@doi [\mnras]
  {10.1111/j.1365-2966.2011.18782.x}, \href
  {http://adsabs.harvard.edu/abs/2011MNRAS.415.1343S} {415, 1343}

\bibitem[\protect\citeauthoryear{{Silk}}{{Silk}}{2018}]{Silk2018Nature}
{Silk} J.,  2018, \mn@doi [\nat] {10.1038/d41586-017-08941-8}, \href
  {https://ui.adsabs.harvard.edu/abs/2018Natur.553....6S} {553, 6}

\bibitem[\protect\citeauthoryear{{Steigman}}{{Steigman}}{2007}]{Steigman2007}
{Steigman} G.,  2007, \mn@doi [Annual Review of Nuclear and Particle Science]
  {10.1146/annurev.nucl.56.080805.140437}, \href
  {http://adsabs.harvard.edu/abs/2007ARNPS..57..463S} {57, 463}

\bibitem[\protect\citeauthoryear{{Sunyaev} \& {Chluba}}{{Sunyaev} \&
  {Chluba}}{2009}]{Sunyaev2009}
{Sunyaev} R.~A.,  {Chluba} J.,  2009, \mn@doi [Astronomische Nachrichten]
  {10.1002/asna.200911237}, \href
  {http://adsabs.harvard.edu/abs/2009AN....330..657S} {330, 657}

\bibitem[\protect\citeauthoryear{{Sunyaev} \& {Khatri}}{{Sunyaev} \&
  {Khatri}}{2013}]{Sunyaev2013}
{Sunyaev} R.~A.,  {Khatri} R.,  2013, \mn@doi [International Journal of Modern
  Physics D] {10.1142/S0218271813300140}, \href
  {http://adsabs.harvard.edu/abs/2013IJMPD..2230014S} {22, 30014}

\bibitem[\protect\citeauthoryear{{Sunyaev} \& {Zeldovich}}{{Sunyaev} \&
  {Zeldovich}}{1970a}]{Sunyaev1970mu}
{Sunyaev} R.~A.,  {Zeldovich} Y.~B.,  1970a, \mn@doi [\apss]
  {10.1007/BF00653472}, \href
  {http://adsabs.harvard.edu/abs/1970Ap%26SS...7...20S} {7, 20}

\bibitem[\protect\citeauthoryear{{Sunyaev} \& {Zeldovich}}{{Sunyaev} \&
  {Zeldovich}}{1970b}]{Sunyaev1970diss}
{Sunyaev} R.~A.,  {Zeldovich} Y.~B.,  1970b, \mn@doi [\apss]
  {10.1007/BF00649577}, \href
  {http://adsabs.harvard.edu/abs/1970Ap%26SS...9..368S} {9, 368}

\bibitem[\protect\citeauthoryear{{Switzer} \& {Hirata}}{{Switzer} \&
  {Hirata}}{2008a}]{Switzer2007I}
{Switzer} E.~R.,  {Hirata} C.~M.,  2008a, \mn@doi [\prd]
  {10.1103/PhysRevD.77.083006}, \href
  {http://adsabs.harvard.edu/abs/2008PhRvD..77h3006S} {77, 083006}

\bibitem[\protect\citeauthoryear{{Switzer} \& {Hirata}}{{Switzer} \&
  {Hirata}}{2008b}]{Switzer2007II}
{Switzer} E.~R.,  {Hirata} C.~M.,  2008b, \mn@doi [\prd]
  {10.1103/PhysRevD.77.083008}, \href
  {http://adsabs.harvard.edu/abs/2008PhRvD..77h3008S} {77, 083008}

\bibitem[\protect\citeauthoryear{{Wong}, {Seager}  \& {Scott}}{{Wong}
  et~al.}{2006}]{Wong2006}
{Wong} W.~Y.,  {Seager} S.,   {Scott} D.,  2006, \mn@doi [\mnras]
  {10.1111/j.1365-2966.2006.10076.x}, \href
  {http://adsabs.harvard.edu/cgi-bin/nph-bib_query?bibcode=2006MNRAS.367.1666W&db_key=AST}
  {367, 1666}

\bibitem[\protect\citeauthoryear{{Zeldovich} \& {Sunyaev}}{{Zeldovich} \&
  {Sunyaev}}{1969}]{Zeldovich1969}
{Zeldovich} Y.~B.,  {Sunyaev} R.~A.,  1969, \mn@doi [\apss]
  {10.1007/BF00661821}, \href
  {http://adsabs.harvard.edu/abs/1969Ap%26SS...4..301Z} {4, 301}

\bibitem[\protect\citeauthoryear{{Zeldovich}, {Kurt}  \&
  {Syunyaev}}{{Zeldovich} et~al.}{1968}]{Zeldovich68}
{Zeldovich} Y.~B.,  {Kurt} V.~G.,   {Syunyaev} R.~A.,  1968, Zhurnal
  Eksperimental noi i Teoreticheskoi Fiziki, \href
  {http://adsabs.harvard.edu/cgi-bin/nph-bib_query?bibcode=1968ZhETF..55..278Z&db_key=AST}
  {55, 278}

\bibitem[\protect\citeauthoryear{{Zeldovich}, {Illarionov}  \&
  {Sunyaev}}{{Zeldovich} et~al.}{1972}]{Zeldovich1972}
{Zeldovich} Y.~B.,  {Illarionov} A.~F.,   {Sunyaev} R.~A.,  1972, SJETP, \href
  {http://adsabs.harvard.edu/abs/1972JETP...35..643Z} {35, 643}

\makeatother
\end{thebibliography}

\bsp	
\label{lastpage}
\end{document}